\newif\ifcqg\cqgtrue
\ifcqg
\documentclass[]{iopart}
\usepackage{iopams}
\usepackage{setstack}
\eqnobysec

\else
\documentclass[aps,letter,prd,thightenlines,nofootinbib,showpacs,eqsecnum,amsfonts,amssymb,amsmath]{revtex4}
\usepackage{dcolumn}
\usepackage{bm}           

\fi

\ifcqg

\else

\fi

\usepackage{epsfig}
\usepackage{dcolumn}
\usepackage{bm}
\DeclareGraphicsExtensions{.eps,.ps,.eps.gz,.ps.gz,.eps.Z,{}}

\newcommand{\rhob}{\overline{\rho}}

\newcommand{\vk}{{\bf k}}
\newcommand{\vq}{{\bf q}}
\newcommand{\vx}{{\bf x}}
\newcommand{\vu}{{\bf u}}
\newcommand{\dd}{{\rm d}}

\newcommand{\mS}{{\cal S}}

\newcommand{\thetar}{\tilde\theta}

\newcommand{\Dirac}{\delta_{\rm D}}
\newcommand{\Phieff}{\Phi^{\rm eff.}}
\newcommand{\Omm}{\Omega_{\rm m}}

\begin{document}
\title{Cosmological Large-scale Structures beyond Linear Theory in  Modified Gravity}

\author{Francis Bernardeau and Philippe Brax}
\ifcqg
\ead{francis.bernardeau@cea.fr, philippe.brax@cea.fr}
\address{CEA, Institut de Physique Th{\'e}orique, 91191 Gif-sur-Yvette c{\'e}dex, France
CNRS, URA-2306, 91191 Gif-sur-Yvette c{\'e}dex, France}
\fi

\vskip 0.15cm

\begin{abstract}
We consider the effect of modified gravity on the growth of large-scale structures at second order in perturbation theory. We show that modified gravity models changing the linear growth rate of fluctuations are also bound to change, although mildly, the mode coupling amplitude  in the density and reduced velocity fields. We present explicit formulae which describe this effect.
We then focus on models of modified gravity involving a scalar field coupled to matter, in particular chameleons and dilatons, where it is shown that there exists a transition scale around which the existence of an extra scalar degree of freedom induces significant changes in the
coupling properties of the cosmic fields. We obtain the amplitude of this effect for realistic dilaton models at the tree-order
level for the bispectrum, finding them to  be comparable in amplitude to those obtained in the DGP model.
\end{abstract}

 \vskip2pc

\ifcqg
\fi
\section{Introduction}

The origin of the acceleration of the expansion of the universe~\cite{Copeland:2006wr,Carroll:2000fy,Peebles:2002gy,Brax:2009ae} is still unknown despite more than ten years of intense experimental and theoretical activities. Four broad types of explanations have been put forward so far. The oldest and also the boldest one posits that the acceleration of the universe is simply due to a cosmological constant~\cite{Weinberg:1988cp}. Of course this approach is fraught with difficulties and clashes with our present understanding of quantum field theory and the structure of quantum corrections. A combination of anthropic arguments and the the existence of a stringy landscape~\cite{Bousso:2000xa} have been used to justify the cosmological constant solution to the acceleration puzzle. Another possibility relies on the plausible existence of large voids in the universe~\cite{Buchert:2007ik}. In a sense, we would have been fooled and the acceleration of the universe would simply be the result of local inhomogeneities. This implies a strong violation of the Copernican principle which still needs to be tested and confirmed (see for instance~\cite{2008PhRvL.100s1303U}).
A more mundane approach following in the footsteps of the early universe inflation era would be to assume that a scalar field leads to the recent acceleration era~\cite{Wetterich:1987fm,Ratra:1987rm,Steinhardt:1999nw}. Such models of dark energy run into three major difficulties. The first one is a fine tuning issue akin to the cosmological constant problem and related to the quantum stability of the scalar potential. Another severe problem is the apparent almost coincidence between the beginning of the matter era and the start of the acceleration phase. This requires some type of tuning which may be avoided using tracking models of quintessence. Finally, a very serious problem arises as the mass of the scalar field now is of the order of the Hubble rate and therefore potentially mediates a long range fifth force~\cite{Will:2001mx}. Although the smallness of the scalar mass is inevitable on large scales, it turns  out that locally, where gravity has been tested, non-linear effects can prevent the observation of a fifth force. This is the case for Galileon models~\cite{Nicolis:2008in} where the Vainshtein mechanism~\cite{Vainshtein:1972sx} is at play, for chameleon models~\cite{Khoury:2003rn,Khoury:2003aq,Brax:2004qh,Mota:2006ed,Mota:2006fz} where a thin shell effect is present and for dilatons~\cite{Gasperini:2001pc} where a variation of the Damour-Polyakov mechanism~\cite{Damour:1994zq} can be implemented~\cite{Damour:2002mi,Damour:2002nv,Brax:2010gi}. Recently a new possibility called the symmetron was presented with similar properties to the Damour-Polyakov mechanism\cite{Hinterbichler:2010es}. In these cases, the acceleration of the universe and the compatibility with local tests of General Relativity is guaranteed. Finally, a fourth type of explanation has been advocated, it involves a modification of gravity. Examples of such models are the DGP model~\cite{Dvali:2000hr} coming from extra dimensional physics and the $f(R)$ models~\cite{Brookfield:2006mq,Faulkner:2006ub,Hu:2007nk,Brax:2008hh}. It turns out that these models of modified gravity are nothing but disguised scalar field models with the Vainshtein and the chameleon mechanisms respectively at play. Hence in the following we will focus on modifications of gravity involving a scalar field. In particular, we will consider the cases of chameleons and dilatons.

Cosmologically the models we are considering differ very little from a pure $\Lambda$-CDM model at least since Big Bang Nucleosynthesis (BBN). The only possibility of detecting modified gravity effects in this context is at the perturbative level. At the linear level and in a first order approximation, it is known that the main feature of these models is the existence of a characteristic scale corresponding to the Compton wave length of the scalar field~\cite{Brax:2004qh,Brax:2009ab}. For scales larger than the Compton wave length, the growth of structure is identical to the one in $\Lambda$-CDM whereas a regime of anomalous growth appears within the Compton scale. We extend these results to second order in perturbation theory~\footnote{In~\cite{2001ApJ...548...47G} one can find an earlier attempt to quantify the effect of modified gravity, in the context of scalar tensor theories, to higher order couplings. The effects of modified gravity were however limited to changes in the background properties.} as a first step towards a more general treatment of the renormalisation group type~\cite{2006PhRvD..73f3519C,2006PhRvD..73f3520C}.

The analysis we carry out applies to general scalar-tensor theories but does not cover other modifications of gravity such as the DGP and Galileon models. In the DGP case, second order perturbation theory has been studied with modifications of General Relativity of similar nature as the ones we will present~\cite{2009PhRvD..80j4006S}. In particular, we find that the perturbation equations contain a time and scale dependent modification of Newton's constant as well as the appearance of extra terms at the second order level. A striking difference between DGP and scalar tensor theories is the fact that for DGP the scale dependence essentially occurs at large scales whereas for scalar-tensor theories this phenomenon is present below the Compton wavelength of the scalar field. We will see however that these classes of models bring much more important changes to the coupling structure than for instance the clustering quintessence model explored in ~\cite{2011arXiv1101.1026S}.

In the following we first introduce the chameleon and dilaton models model we are interested in,  their motivations and basic properties, in section \ref{ModGravMod}. In section \ref{gammaModel} we present the generic properties of second order calculations and how modification of gravity can affect the mode coupling amplitudes. The sections \ref{LinearEffects} and \ref{NonLinearEffects} explore in more details the consequences of a realistic change of gravity, either due to  a change of the linear growth rate or the presence of  extra coupling terms. We finally conclude and discuss our results in view of future experiments.

\section{Scalar Fields and Modified Gravity}
\label{ModGravMod}

We are mainly interested here in models of modified gravity involving a scalar field. This is the case of the Galileon model, f(R) theories, chameleonic and dilatonic models. The last three types will be briefly analysed in this section.

\subsection{Chameleon models}

Dark energy models suffer from the usual smallness of the scalar field mass on cosmological scales. Chameleon models overcome this problem by introducing a coupling of the scalar field to matter leading to a density dependent mass. In a dense environment chameleons have a large mass which leads to the absence of gravity violation in the solar system and the laboratory.
Cosmologically, the chameleon models can be viewed as
modified gravity models with a constant coupling  to matter $\beta$ and a scalar field with a time-varying mass $m(a)$. In these models, the bare potential appearing in the Lagrangian is modified by the interaction with the environment.
The action governing the dynamics of the chameleon field $\phi$  is given by
\begin{eqnarray}
S&=&\int d^4x\sqrt{-g}\left\{\frac{M_{\rm Pl}^2}{2}{\cal
R}-\frac{1}{2}g^{\mu \nu} \partial_\mu \phi \partial_\nu \phi- V(\phi)\right\}\nonumber\\
&& +
 \int d^4x \sqrt{-\tilde{g}} {\cal
L}_m(\psi_m^{(i)},\tilde g_{\mu\nu})\,, \label{action}
\end{eqnarray}
where $g_{\mu \nu}$ is the metric in the Einstein frame, $\tilde g_{\mu\nu}$ is the metric in the Jordan frame,
$M_{\rm Pl}\equiv (8\pi G_N)^{-1/2}\equiv {\kappa_4^{-1}}$ is the reduced Planck mass, ${\cal R}$ is
the Ricci scalar for the Einstein-frame metric, and $\psi_m^{(i)}$ are various matter fields
labeled by $i$.
The Einstein-frame metric $g_{\mu \nu}$ and the Jordan-frame metric $\tilde g_{\mu\nu}$ are related
by the conformal rescaling
\begin{equation}
\tilde g_{\mu\nu}=e^{2\beta\phi/M_{\rm Pl}}g_{\mu\nu}\, \equiv A^2(\phi) g_{\mu\nu},
\label{conformal}
\end{equation}
where $\beta$ is a  dimensionless constant. In chameleon models this constant
is assumed to be the same for all types of matter, respecting the weak equivalence principle where
all material particles feel the same metric.

The Klein-Gordon  equation in the presence of non-relativistic matter defines the following effective potential for the scalar field
\begin{equation}
V_{\rm eff}(\phi) =   V(\phi) +  {\rho} e^{\beta\phi/M_{\rm Pl}}.
\label{veff}
\end{equation}
where $\rho$ is the conserved matter density in the Einstein frame.
It follows 
that $V_{\rm eff}$ has a minimum at $\phi_{\rm min}({\rho})$ which is solution to
\begin{equation}
\frac{dV}{d\phi}= -\frac{\beta}{M_{\rm Pl}}e^{\beta\phi/M_{\rm Pl}} {\rho}
\label{minphi}
\end{equation}
when the bare potential is of the runaway type as in freezing models of quintessence.
The mass at the minimum of the effective potential is given by
\begin{equation}
m^2= \frac{d^2 V}{d\phi^2}+ \frac{\beta^2}{M_{\rm Pl}^2}e^{\beta\phi/M_{\rm Pl}} {\rho}.
\label{msqdef}
\end{equation}
As $\frac{dV}{d\phi}<0$ and $\frac{d^2V}{d\phi^2}>0$ for successful chameleon models, we find that
\begin{equation}
m^2\ge 3 \beta^2 H^2\,,
\end{equation}
where we have assumed that the chameleon energy density is negligible compared to ${\rho}$, $\rho_\phi \ll \rho$.
In contrast to usual quintessence models, in most cases the mass of the chameleon
is much larger than the Hubble rate  and up to numerical factors of order unity,
\begin{equation}
m^2 \simeq
3 \beta H^2 \frac{M_{\rm Pl}}{\phi_{\rm min}(\rho)} \gg H^2  \qquad {\rm when } \qquad
\phi\ll M_{\rm Pl}.
\label{HH}
\end{equation}
Thus it is possible to have fluctuations in the chameleon field at scales much smaller than
the horizon scale. This is of cosmological importance for structure formation.
When Eq. (\ref{HH}) is satisfied, the minimum of the effective potential is an attractor,
and the chameleon therefore settles down there, and evolves slowly since at least Big Bang Nucleosynthesis.
In that case, the potential energy of the chameleon is nearly constant and the model is equivalent to a $\Lambda$-CDM (Cold Dark Matter)  model.

The dynamics of chameleon models are not equivalent to
a $\Lambda$-CDM model at the level of perturbations. The first order perturbation theory is well-known with an anomalous growth of structures for scales $k/a \ll m(a)$, i.e. within the Compton wavelength of the chameleon. We will extend this result to second order shortly.

Another class of equivalent scalar-tensor theories is given by the so-called $f(R)$ models with a Lagrangian
\begin{equation}
S=\frac{1}{2\kappa_4^2}\int d^4 x {\sqrt -g} f(R).
\end{equation}
In fact these theories are equivalent to  scalar field models with a scalar field $\phi_R$ identified as
\begin{equation}
\frac{d f}{dR}= e^{-2\beta \kappa_4 \phi_R}
\end{equation}
and $\beta=1/\sqrt{6}$. The potential of these theories is given by
\begin{equation}
V(\phi_R)= m_{\rm Pl}^2 \frac{ R f_R -f}{2f_R^2}
\end{equation}
and the coupling function is
\begin{equation}
A(\phi_R)= e^{\beta \kappa_4 \phi_R}.
\end{equation}
In this guise, we see that $f(R)$ models must behave like chameleon fields in order to evade gravitational constraints\cite{Brax:2008hh}. These models essentially behave like $\Lambda$-CDM in the recent past of the universe. Their perturbative properties are crucial to try to distinguish them from $\Lambda$-CDM.

In the following we will focus on chameleon models defined by an inverse power law potential
\begin{equation}
V(\phi)= \Lambda^4 + \frac{\Lambda^{4+n}}{\phi^n}+ \dots
\end{equation}
where we neglect higher inverse powers of the chameleon field.
The effective potential has a minimum at
\begin{equation}
\phi_{\rm min}=\left(\frac{n M_{\rm Pl} \Lambda ^{4+n}}{\beta
\rho} \right)^{1/(n+1)}.
\end{equation}
and the mass of the field at the minimum is given in Eq.~(\ref{msqdef}).
For this model (assuming $\phi \ll M_{\rm Pl}$),
\begin{equation}
m^2= (n+1) \beta \left( \frac{\rho}{M_{\rm Pl}^2} \right) \left( \frac{M_{\rm Pl}}{\phi_{\rm min}}\right) \gg (n+1) \beta \left( \frac{\rho}{M_{\rm Pl}^2} \right).
\label{mmin}
\end{equation}
In a cosmological setting, the Hubble parameter $H \simeq \sqrt{\rho}/M_{\rm Pl}$ so that $m\gg \sqrt{\beta} H$.
Hence, analogously to the previous case, the field quickly relaxes to the minimum of the potential and
sits there for most of the cosmological history, at least since Big Bang Nucleosynthesis (BBN).
The fact that the scalar field can have a mass much larger than $H$ is one of the main differences between
chameleon models and usual quintessence models.
At the minimum of the potential,
\begin{equation}
V_{\rm eff}(\phi_{\rm min}) \approx \Lambda^4 + \rho.
\end{equation}
This implies that the model behaves like a pure cosmological constant in the recent past. The matter contribution is simply the usual energy density entering the Friedmann equation.

Using the fact that these models behave like $\Lambda$-CDM, we have during the matter epoch $\rho = \rho_{\rm eq} ({a_{\rm eq}}/{a})^{3}$. This implies that the field and the mass have a time dependence given by,
\begin{equation}
\phi_{\rm min}\sim a^{3/(n+1)}
\end{equation}
and
\begin{equation}
m^2(a) \sim a^{-3(n+2)/(n+1)}.
\end{equation}
It is important to  notice that the comoving Compton wavelength  scales like
\begin{equation}
m(a) a \sim a^{-(n+4)/2(n+1)}
\end{equation}
which decreases with time. Hence a given scale $k$ enters the Compton wavelength at late time implying that gravity is essentially modified
at short distance in the recent past of the universe.

\subsection{General scalar-tensor theories and dilaton models}

So far we have focused on models where the mass of the scalar field is time-dependent and the coupling $\beta$ is constant. More general models where the coupling is a field function itself can be easily constructed.
In the following we consider  models with the following action
\begin{eqnarray}
S&=&\int d^4x\sqrt{-g}\left\{\frac{M_{\rm Pl}^2}{2}{\cal
R}-{M_{\rm Pl}^2 }g^{\mu \nu}  k^2(\phi) \partial_\mu \phi \partial_\nu \phi- V(\phi)\right\} \nonumber\\
&&+ \int d^4x \sqrt{-\tilde{g}} {\cal
L}_m(\psi_m^{(i)},A^2(\phi) g_{\mu\nu})\,, \label{action}
\end{eqnarray}
Notice that $k^2(\phi)=1/2M_{\rm Pl}^2$ leads to a canonically normalised field.
A particularly interesting case concerns the runaway dilaton of string cosmology in the strong coupling regime of string theory\cite{Gasperini:2001pc}.
In this dilatonic case we have
\begin{equation}
V(\phi)= A^4(\phi) V_0e^{-\phi}
\end{equation}
in the large $\phi$ limit corresponding to the large string coupling limit and
\begin{equation}
k^2(\phi)= 3\beta^2(\phi) - A^2(\phi)\frac{Z(\phi)}{2c_1^2}
\end{equation}
where
\begin{equation}
\beta(\phi)=\frac{d\ln A}{d\phi}
\end{equation}
is the field dependent coupling constant,
and we have
\begin{equation}
Z(\phi)= -\frac{2c_1^2}{\lambda^2}+ b_Z e^{-\phi} +\dots
\end{equation}
In the strong coupling limit of string theory, Damour and Polyakov in \cite{Damour:1994zq} have argued that the coupling should vanish at a finite/infinite value
of $\phi$:
\begin{equation}
A(\phi)= 1 +\frac{A_2}{2} (\phi-\phi_0)^2 +\dots\label{Amodel}
\end{equation}
which has a minimum at $\phi_0$. The coefficient $A_2\sim M_{\rm Pl}^2/M_{s}^2$ where $M_s$ is the string scale must be large enough  to satisfy the solar system tests\cite{Brax:2010gi}.

In these models
the effective potential is
\begin{equation}
V_{\rm eff}(\phi)= V(\phi) + A(\phi) \rho_m
\end{equation}
where $\rho_m$ is the conserved matter density in the Einstein frame.
We  define the  normalised field
\begin{equation}
d\varphi= k(\phi) d\phi
\end{equation}
whose mass is
\begin{equation}
m^2(\bar \phi)=4\pi G\ \frac{d^2 V_{\rm eff}}{d\varphi^2}
\end{equation}
where $\bar\phi$ is the cosmological background value. At the background level, like chameleon models, the dilaton models behave like a $\Lambda$-CDM model. At the perturbative level, we can  define a critical wave mode $k_{c}$
\begin{equation}
k_{c}\equiv a(\eta_{0})m(\bar \phi(\eta_{0}))
\end{equation}
below which gravity is modified now.

\begin{figure}[htbp]
\epsfig{file=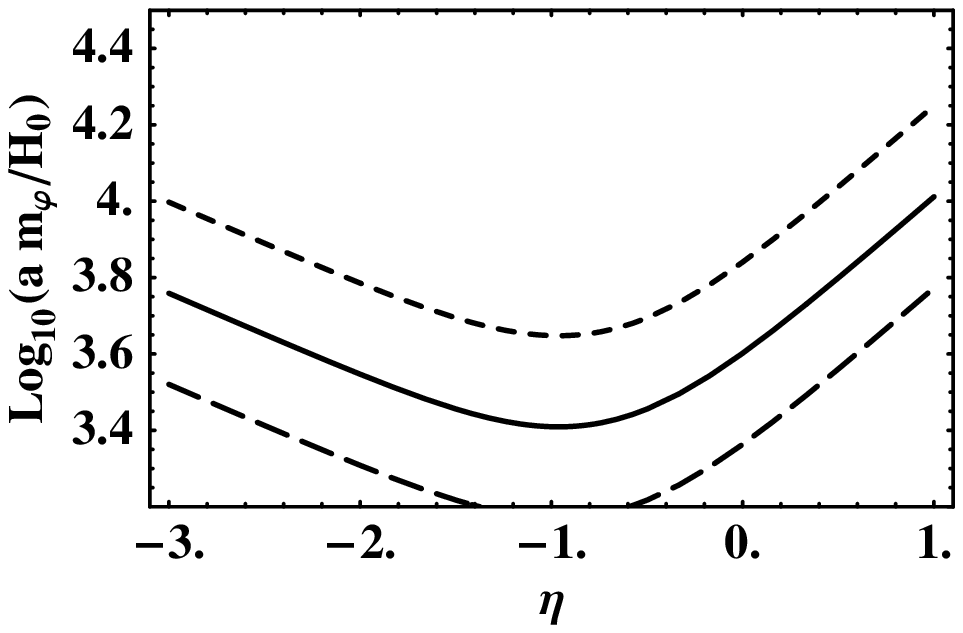,width=4cm}\hspace{.1cm}
\epsfig{file=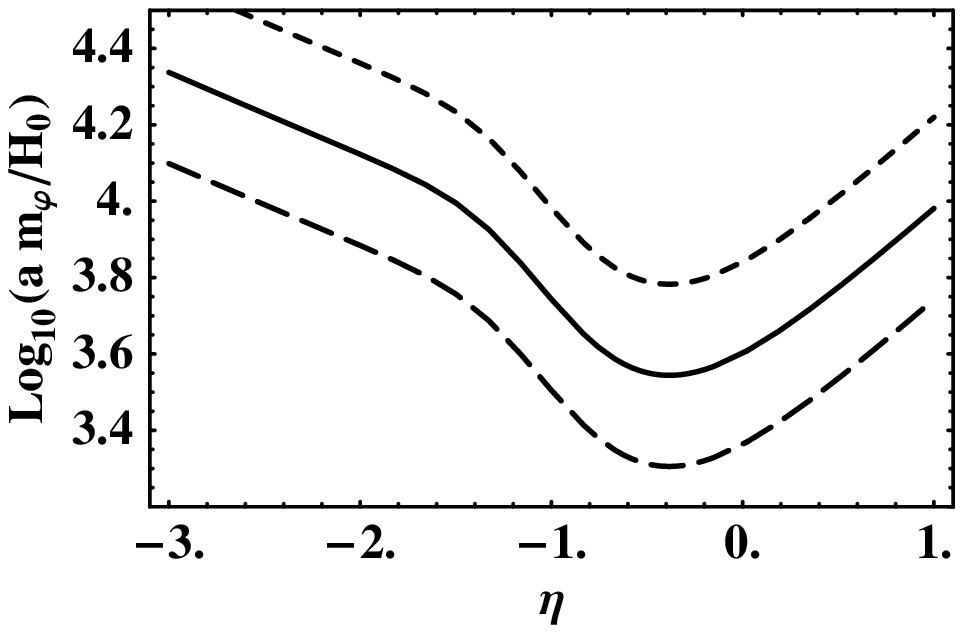,width=4cm}\hspace{.1cm}
\epsfig{file=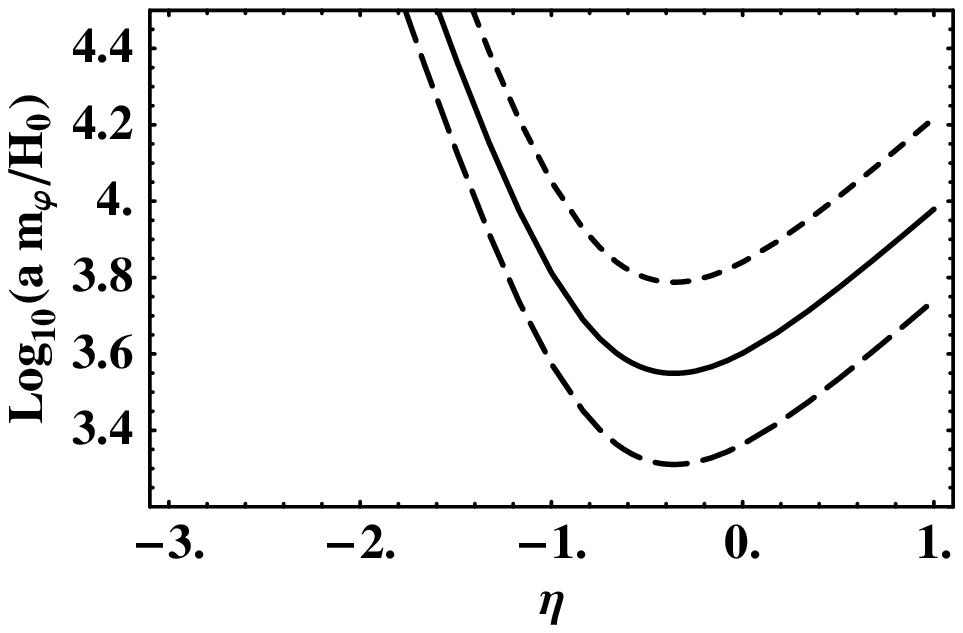,width=4cm}
\caption{The effective mass of the field $\varphi$ as a function of time $\eta$ in units of $H_{0}$. The numerical application corresponds to the model (\ref{Amodel}) with  $\lambda=1$ (left panel), $\lambda=10$ (middle panel) and $\lambda=1000$ (right panel),$A_{2}=5.6\,10^5$ (solid line) and the dashed lines to cases where $A_{2}$ is $1/3$ or $3$ times larger. In this case the value of $k_{c}$ is $k_{c}=4\,10^{3} H_{0}/c \approx 1.3 h Mpc^{-1}$. Gravity is modified for modes that are above the solid line at time $\eta$.}
\label{MassEvol}
\end{figure}

\section{Parametrised Modified Gravity}
\label{gammaModel}

In this section we consider a simple parametrisation of modified gravity and  explore the consequences of this very phenomenological approach. It has been argued in the literature that models of modified gravity could be characterized by the growth rate of the perturbations. We envisage here the phenomenological consequences of such an effect on the higher order correlation functions assuming the change of growth rate is entirely due to a (time dependent) modification of the gravitational force.

\subsection{The $\gamma-$model}

As a first approximation before considering more general theories, we focus on models where matter is conserved:
\begin{equation}
\frac{\partial \delta(\vx,t)}{\partial t}+\frac{1}{a}[(1+\delta)\vu_{i}(\vx,t)]_{,i}=0.
\end{equation}
and  $\delta(\vx,t)$ is the density contrast at comoving position $\vx$ and physical time $t$. This equation is valid on sub-horizon
scales. Here and in the following we do not make the distinction between the dark matter fluid and the baryon fluid.
Then in the single flow approximation, neglecting pressure terms,  the Euler equation takes the form,
\begin{eqnarray}
\hspace{-1cm}\frac{\partial \vu_{i}(\vx,t)}{\partial t}+H(t) \vu_{i}(\vx,t)+\frac{1}{a(t)}\vu_{j}(\vx,t)\vu_{i,j}(\vx,t)=-\frac{1}{a(t)}\,\Phieff_{,i}(\vx,t)
\end{eqnarray}
where $\Phieff(\vx,t)$ is the effective potential seen by  the massive particles. Before exploring more realistic settings, we simply assume here that it is given by
\begin{equation}
\Phieff(\vx,t)=(1+\epsilon(t))\Phi(\vx,t)
\end{equation}
where $\Phi(\vx,t)$ is the standard Newtonian potential; i.e. it satisfies the standard Poisson equation,
\begin{equation}
\frac{1}{a^2(t)}\Delta\Phi(\vx,t)=4\pi G\rhob(t)\,\delta(\vx,t)\label{PoissonEq}.
\end{equation}
One can then  introduce $\theta=\vu_{i,i}/(aH)$ which is nothing but the local expansion rate in units of $H$.
Then the conservation equations can be written in Fourier space,
\begin{eqnarray}
\hspace{-1cm}\dot\delta(k,t)+H\theta(k,t)=-H\,\Dirac(\vk-\vk_{1}-\vk_{2})\,\alpha(\vk_{1},\vk_{2})\delta(k_{1},t)\ \theta(k_{2},t)\label{FourierCont}\\
\hspace{-1cm}\dot\theta(k,t)+\left(2H+\frac{\dot H}{H}\right)\theta(k,t)+\frac{3}{2}H^2\Omm(1+\epsilon(t))\delta(k,t)=
\nonumber\\
\hspace{2cm}-H\,\Dirac(\vk-\vk_{1}-\vk_{2})\,
\beta(\vk_{1},\vk_{2})\theta(k_{1},t)\theta(k_{2},t)\label{FourierEul}
\end{eqnarray}
where $\alpha(\vk_{1},\vk_{2})=\vk_{2}.(\vk_{1}+\vk_{2})/k_{2}^2$ and
$\beta(\vk_{1},\vk_{2})=(\vk_{1}.\vk_{2})(\vk_{2}.(\vk_{1}+\vk_{2}))/k_{1}^2/k_{2}^2$ and
where a dot stands for a derivative with respect to time. The right hand side terms are implicitly assumed to
be integrated over $\vk_{1}$ and $\vk_{2}$.
We have furthermore introduced the density parameter $\Omm(t)$ which represents the (pressureless) matter density in units of the critical density.

When linearized this system leads to a scale independent linear system for the growth rate $D$,
\begin{equation}
\ddot D(t)+2H\dot D(t)-\frac{3}{2}H^2\Omm(1+\epsilon(t))D(t)=0\label{lingrowth}
\end{equation}
Let us denote by $D_{+}$ the solution of this equation which grows with time (assuming there is only one of such solutions). This is the case for standard gravity, for which $\epsilon=0$, so by continuity it should be similar  when $\epsilon$ is small enough. We can then define the factor $f(t)$ as
\begin{equation}
f(t)=\frac{\dd \log D_{+}(k,t)}{\dd \log a(t)}.\label{fdef}
\end{equation}
so that the time dependence of $\theta(k,t)$ for the growing mode is nothing but $f(t)\,D_{+}(t)$.

In this section, and  in this section only, we assume  that $f(t)$ takes the form,
\begin{equation}
f(t)=\Omm^{\gamma},\label{fgamform}
\end{equation}
and that $\epsilon(t)$ is such that $f(t)$ has such an expression when $\Omm(t)$ is given by the background evolution of the standard
$\Lambda-$CDM model. It is known (and we will recover this result) that for standard gravity
\begin{equation}
\gamma^{\rm GR}\approx 0.55.
\end{equation}
For such a standard background evolution we have $\dot{H}/H^2=-3/2\,\Omm$ and $\dot\Omm=3H\Omm(\Omm-1)$
so that one can encapsulate the entire time dependence in the $\Omm(t)$ dependence.

Assuming  Eq. (\ref{fgamform}),  the linearized conservation equations can be used to infer
the form of $\epsilon(t)$:
\begin{equation}
1+\epsilon_{\gamma}(t)=\frac{1}{3\Omm/2}\left[\Omm^{2\gamma}+3\gamma\Omm^{\gamma}(\Omm-1)+(2-3\Omm/2)\Omm^{\gamma}\right].\label{epsgam}
\end{equation}
In the next paragraph we will analyse the consequences of such a change for the growth of higher order correlation functions.

\subsection{Bispectra in the $\gamma-$model}

The bispectra will be driven, at large enough scales, by the mode coupling evolution deduced from  the equations of motion.
Let us assume that one can expand the cosmic fields with respect to the initial (linear) fields,
\begin{equation}
\delta(\vk)=\sum_{p}\delta^{(p)}(\vk),\ \ \theta(\vk)=\sum_{p}\theta^{(p)}(\vk).
\end{equation}
where $\delta^{(p)}(\vk)$ is of order $p$ in   $\delta^{(1)}(\vk)$. Generically we can write,
\begin{equation}
\delta^{(p)}(\vk)\sim F_{p}\left[\delta^{(1)}(\vk)\right]^{p}
\end{equation}
and similarly for the velocity divergence with functions $G_{p}$ (we will make these forms more precise in the following).
Then for Gaussian initial conditions the tree-order density bispectra are directly related to the
second order density field. More precisely if we define the density spectrum $P(k)$ and bispectrum $B(k_{1},k_{2},k_{3})$ as
\begin{eqnarray}
\langle\delta(\vk_{1})\delta(\vk_{2})\rangle&=&(2\pi)^3\delta_{D}(\vk_{1}+\vk_{2})P(k_{1})\\
\langle\delta(\vk_{1})\delta(\vk_{2})\delta(\vk_{3})\rangle&=&(2\pi)^3\delta_{D}(\vk_{1}+\vk_{2}+\vk_{3})B_{\delta}(k_{1},k_{2},k_{3})
\end{eqnarray}
then at tree order
\begin{equation}
B_{\delta}(k_{1},k_{2},k_{3})=2F_{2}(\vk_{1},\vk_{2})P(k_{1})P(k_{2})+\hbox{sym.}
\end{equation}
Note that for convenience one usually defines the reduced bispectrum as
\begin{equation}
Q_{\delta}(k_{1},k_{2},k_{3})=\frac{B_{\delta}(k_{1},k_{2},k_{3})}{P(k_{1})P(k_{2})+P(k_{2})P(k_{3})+P(k_{3})P(k_{1})}
\end{equation}
which is then independent of the amplitude of the spectrum.

The quantities that are accessible to observations are generically the density field, for instance the projected density
field in case of the cosmic shear, or, for redshift space observations, a combination of the density field and the \textsl{reduced}
velocity divergence, where the reduced velocity divergence $\thetar(\vx,t)$ is defined by $\thetar(\vx,t)=\theta(\vx,t)/f(t)$.
In the following we simply assume that the bispectrum $Q_{\delta}$ or similarly that of the reduced divergence $Q_{\theta}$
are both accessible to observations and we will therefore analyse the impact  of modified gravity models on these
quantities.

\begin{figure}[htbp]
\epsfig{file=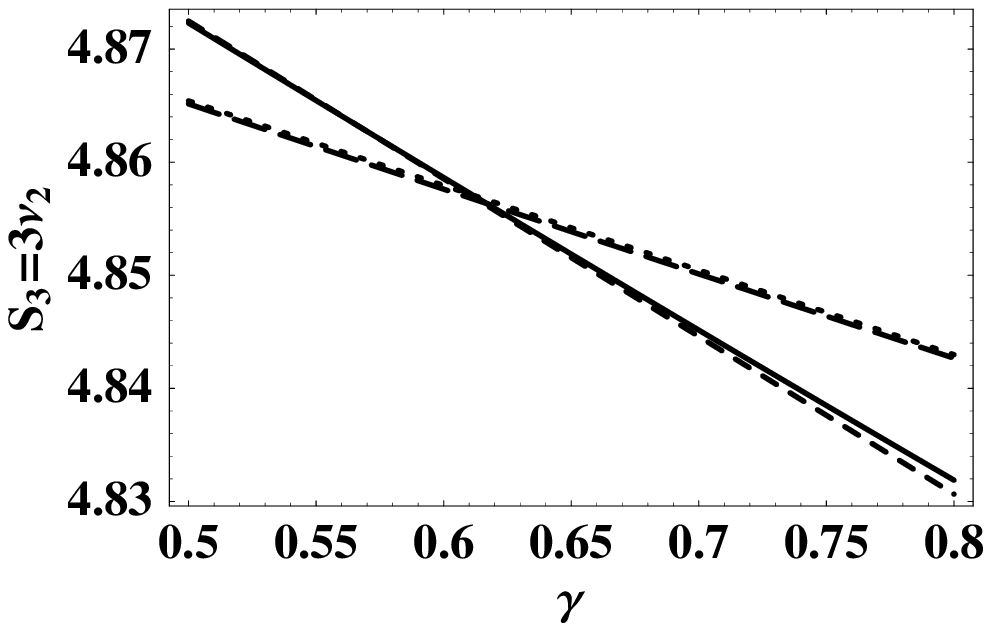,scale=0.5}\hspace{.3cm}\epsfig{file=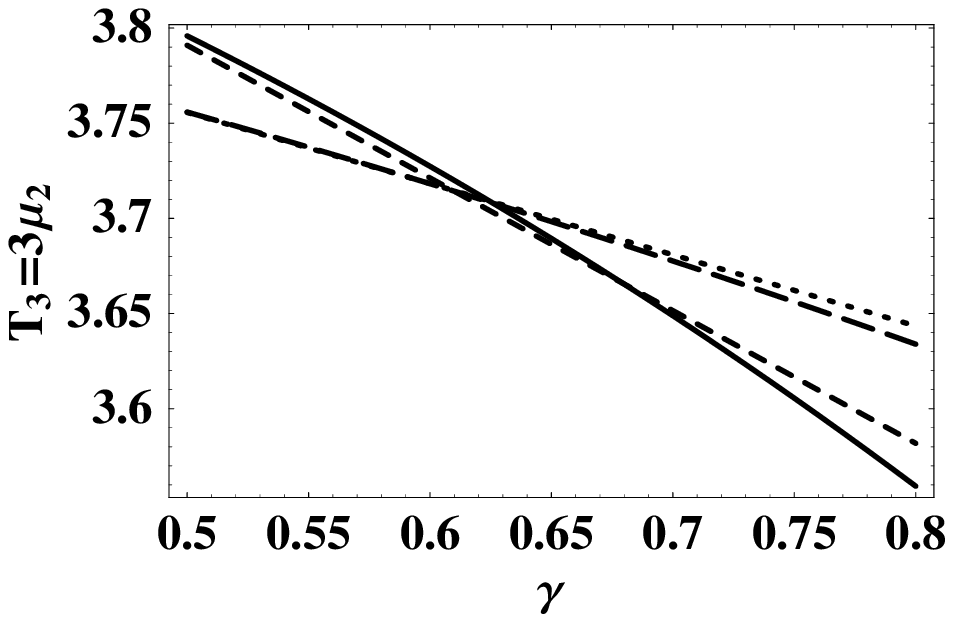,scale=0.5}
\caption{Value of $\nu_{2}$ (left panel) and $\mu_{2}$ (right panel) as a function of $\gamma$ in the $\gamma-$model for $z=0$ (solid line) and $z=1$ (dashed line) and the expressions (\ref{nu2fit}-\ref{mu2fit}) (corresponding dotted lines). }
\label{S3Pream}
\end{figure}

In Fourier space the equations of motion then become,
\begin{eqnarray}
\delta'(k)+{\thetar}(k)=-\Dirac(\vk-\vk_{1}-\vk_{2})\alpha(\vk_{1},\vk_{2})\,\delta(k_{1})\,\thetar(k_{2})\label{PrContExp}\\
\thetar'(k)-\left(1-\frac{3}{2}\frac{\Omm}{f^2}(1+\epsilon)\right)\thetar(k)+\frac{3}{2}\frac{\Omm}{f^2}(1+\epsilon)\delta(k)=\nonumber\\
\hspace{3cm}-\Dirac(\vk-\vk_{1}-\vk_{2})\beta(\vk_{1},\vk_{2})\thetar(k_{1})\thetar(k_{2})\label{PrEulerExp}
\end{eqnarray}
where a prime denotes a derivation with respect to the logarithm of the growing mode,
\begin{equation}
X' \equiv D_{+}\frac{\partial X}{\partial D_{+}}=\frac{1}{f(t)\,H}\dot X.
\end{equation}
The system (\ref{PrContExp}-\ref{PrEulerExp}) can be solved recursively when one assumes that the fields
$\delta$ and $\thetar$ can be expanded with respect to the linear solution,
\begin{equation}
\delta(\vk)=\sum_{p}\delta^{(p)}(\vk),\ \ \thetar(\vk)=\sum_{p}\thetar^{(p)}(\vk).
\end{equation}
The leading time dependence of $\delta^{(p)}$ (and $\thetar^{(p)}$) can easily be deduced and goes like $D_{+}^{p}$. On can then formally define the functions $F_{p}$ and $G_{p}$ as
\begin{eqnarray}
\delta^{(p)}(\vk)&=&\int\dd^3\vk_{1}\dots\dd^3\vk_{p}\,\delta_{\rm Dirac}(\vk-\sum_{i}\vk_{i})\nonumber\\
&&\hspace{1cm}\times F_{p}(\vk_{1},\dots,\vk_{p})\delta^{(1)}(\vk_{1})\dots\delta^{(1)}(\vk_{p})\label{dpexp}\\
\thetar^{(p)}(\vk)&=&-\int\dd^3\vk_{1}\dots\dd^3\vk_{p}\,\delta_{\rm Dirac}(\vk-\sum_{i}\vk_{i})\nonumber\\
&&\hspace{1cm}\times G_{p}(\vk_{1},\dots,\vk_{p})\delta^{(1)}(\vk_{1})\dots\delta^{(1)}(\vk_{p}).\label{tpexp}
\end{eqnarray}
In general this system can be solved when $\xi(t)$ defined by,
\begin{equation}
\xi(t)=\frac{\Omm}{f^2}(1+\epsilon),
\end{equation}
is independent of time.
The functions $F_{p}$ and $G_{p}$ are then independent of time and the system (\ref{PrContExp}-\ref{PrEulerExp}) gives recursive relations (this extends the relations (43-44) of the review paper~\cite{2002PhR...367....1B} obtained for an Einstein-de Sitter background in GR) for the functions $F_{p}$ and $G_{p}$,
\begin{eqnarray}
pF_{p}(\vk_{1},\dots,\vk_{p})-G_{p}(\vk_{1},\dots,\vk_{p})=\nonumber\\
\hspace{2cm}\sum_{q=1}^{p-1}\alpha(\vq_{1},\vq_{2})F_{p-q}(\vk_{q+1},\dots,\vk_{p})G_{q}(\vk_{1},\dots,\vk_{q})\\
\left(1-p-\frac{3\xi}{2}\right)G_{p}(\vk_{1},\dots,\vk_{p})+\frac{3\xi}{2}F_{p}(\vk_{1},\dots,\vk_{p})=\nonumber\\
\hspace{2cm}-\sum_{q=1}^{p-1}\beta(\vq_{1},\vq_{2})G_{p-q}(\vk_{q+1},\dots,\vk_{p})G_{q}(\vk_{1},\dots,\vk_{q})
\end{eqnarray}
with $F_{1}=G_{1}=1$, $\vq_{1}=\sum_{i=1}^{q}\vk_{i}$, $\vq_{2}=\sum_{i=q+1}^{p}\vk_{i}$.
When  the coefficient $\xi$ is time dependent the situation is slightly more complicated but the general forms of the functions $F_{2}$ and $G_{2}$ is still very constrained.
From the continuity equation it is easy to see that we have the relation,
\begin{equation}
G_2(\vk_1,\vk_2)= 2 F_2 (\vk_1,\vk_2)-\frac{\alpha (\vk_1,\vk_2)+\alpha( \vk_2,\vk_1)}{2}
\end{equation}
It is slightly  more cumbersome to see that at late time the functions $F_{2}$ and $G_{2}$ necessarily take the form\footnote{This comes from the decomposition of the right hand side of Eqs. into the growing and decaying modes of the linear system. It turns out that
${\vk_{1}.\vk_{2}}/{k_{1}^2}+{\vk_{1}.\vk_{2}}/{k_{2}^2}$ is always in the growing mode.},
\begin{equation}
\hspace{-1cm}F_{2}(\vk_{1},\vk_{2})=\left(\frac{3\nu_{2}}{4}-\frac{1}{2}\right)+\frac{1}{2}\frac{\vk_{1}.\vk_{2}}{k_{1}^2}+\frac{1}{2}\frac{\vk_{1}.\vk_{2}}{k_{2}^2}
+\left(\frac{3}{2}-\frac{3\nu_{2}}{4}\right)\frac{(\vk_{1}.\vk_{2})^2}{k_{1}^2 k_{2}^2}\label{PrF2Exp}
\end{equation}
and
\begin{equation}
\hspace{-1cm}G_{2}(\vk_{1},\vk_{2})=\left(\frac{3\mu_{2}}{4}-\frac{1}{2}\right)+\frac{1}{2}\frac{\vk_{1}.\vk_{2}}{k_{1}^2}+\frac{1}{2}\frac{\vk_{1}.\vk_{2}}{k_{2}^2}
+\left(\frac{3}{2}-\frac{3\mu_{2}}{4}\right)\frac{(\vk_{1}.\vk_{2})^2}{k_{1}^2 k_{2}^2}\label{PrG2Exp}
\end{equation}
where $\nu_{2}$ and $\mu_{2}$ are the only time dependent coefficients~\footnote{Actually $\nu_{2}$ and $\mu_{2}$ can be interpreted as the geometrical average of $F_{2}$ and $G_{2}$ respectively and more generally the background dependence of $F_{p}$ and
$G_{p}$ are entirely encoded into their geometrical averages which in turns can be derived from the spherical collapse dynamics. See section 2.4 of \cite{2002PhR...367....1B} for details.} which depend on the background through ${\Omm}(1+\epsilon)/f^2$.
For Einstein-de Sitter in GR, this is simply unity and  it is easy to find that $\nu_{2}=34/21$ and $\mu_{2}=26/21$.
The expressions of $\nu_{2}$ and $\mu_{2}$ can easily be computed numerically for the $\gamma-$model.
The results are presented on Figs. \ref{S3Pream}.

In the standard General Relativity (GR) regime and for a matter dominated universe, $\xi\approx 1$ so that the results must be close to the Einstein-de Sitter case at early time (high redshift) and gradually depart from it at late time. To a large extent, the results obtained here can actually be derived from a perturbation calculation when $\Omega\approx 1$.
Let us assume that
\begin{equation}
\eta_\Omega =1-\Omm
\end{equation}
is a small (time) dependent variable.
Then from Eq. (\ref{epsgam}) one has,
\begin{equation}
\epsilon_{\gamma}\approx2\left(1-\frac{11}{6}\gamma\right)\eta_\Omega.
\end{equation}
The parameters $\nu_{2}$ and $\mu_{2}$ can similarly be computed perturbatively about the Einstein-de Sitter solution,
\begin{equation}
\nu_{2}=\frac{34}{21}\left(1+\eta_{\nu}\right),\ \ \mu_{2}=\frac{26}{21}\left(1+\eta_{\mu}\right)
\end{equation}
and one finds
\begin{equation}
\eta_{\nu}=\frac{1}{221}(3-5 \gamma)\eta_\Omega,\ \ \eta_{\mu}=\frac{5}{169}(3-5\gamma)\eta_\Omega.
\end{equation}
Note that in this approach the GR result corresponds to $\gamma=6/11$ (indeed close to 0.55) for which $\epsilon_{\gamma}$ vanishes.
In this case we actually find that the $\nu_{2}$ and $\mu_{2}$ parameters are well parametrised\footnote{This form for $\nu_{2}$ is compatible with the expansion we found and suggested by the expression of the second order displacement field as shown in \cite{2002PhR...367....1B}. The expression for $\mu_{2}$ in this reference is however incorrect.} by
\begin{eqnarray}
\nu_{2}^{\rm GR}&=&\frac{4}{3}+\frac{2}{7}\Omm^{-1/143}.\\
\mu_{2}^{\rm GR}&=&-\frac{4}{21}+\frac{10}{7}\Omm^{-1/143}.
\end{eqnarray}
More generally we  find that the relations
\begin{eqnarray}
\nu_{2}^{\rm MG}&=&\nu_{2}^{GR}-\frac{10}{273}(\gamma-\gamma^{\rm GR})(1-\Omm)\Omm^{\gamma^{\rm GR}-1}\label{nu2fit}\\
\mu_{2}^{\rm MG}&=&\mu_{2}^{GR}-\frac{50}{273}(\gamma-\gamma^{\rm GR})(1-\Omm)\Omm^{\gamma^{\rm GR}-1}\label{mu2fit}
\end{eqnarray}
give a good fit to the numerical results for realistic values of $\Omm$.

\section{Linear Scalar-Tensor theories}
\label{LinearEffects}

In this section we will deal with models where the coupling $\beta$ is constant and the potential is the one of a massive particle with a  time dependent mass. This is a first order approximation to the behaviour of chameleon and dilaton models. The effects of the perturbations of the mass term and the coupling function will be dealt with later.

We need to amend the perturbation equations to take into account the effects of modified gravity. We first consider models where matter is conserved and matter particles interact with a new scalar force of constant coupling and a time  varying range. The continuity equation is therefore left unchanged. This simply expresses that matter (here either baryonic or dark matter) is assumed to be conserved during the cosmic evolution.
In the single flow approximation, neglecting pressure terms, but in the presence of an extra scalar field, the Euler equation takes the form,
\begin{equation}
\hspace{-2cm}
\frac{\partial \vu_{i}(\vx,t)}{\partial t}+H(t) \vu_{i}(\vx,t)+\frac{1}{a(t)}\vu_{j}(\vx,t)\vu_{i,j}(\vx,t)=-\frac{1}{a(t)}\left(\Phi(\vx,t)+\beta\,\phi(\vx,t)\right)_{,i}
\end{equation}
where $\Phi$ is the usual Newton potential (the one that enters in the metric) and $\phi(\vx,t)$ an extra scalar field
that couples to the matter field. Both fields are coupled to the matter fluctuations.
The potential $\Phi$ follows the usual Poisson equation Eq. (\ref{PoissonEq}), whereas the $\phi$ field obeys the following equation
\begin{equation}
\frac{1}{a^2(t)}\Delta\phi(\vx,t)-m^{2}\phi=8\pi G\beta \rhob(t)\,\delta(\vx,t).
\end{equation}
where $\beta$ is a free coupling and $\Delta$ is the flat space Laplacian. It determines the strength of the modification of gravity. The other relevant parameter is the mass scale $m$. This mass scale is a priori time dependent\footnote{In the numerical applications we do here we assume that
$
m\,a\sim a^{-3/2}.
$
which corresponds to $n=1/2$ in terms of chameleon models.}.
We will see later that the fact that $m\,a$ is decreasing  with time implies that, for any scale,  gravity was always standard in the past.

This system of equations
form a closed system between the local density contrast, the peculiar velocity $\vu$, the gravitational potential $\Phi$ and $\phi$. Note that these equations are valid in the sub-horizon limit, at linear order in $\phi$ (higher order corrections will be dealt in the following section on general scalar-tensor theories) and  any order in the density contrast.
As for the standard equations of motions, we can note that the source term of the Euler equation is a potential term. As a result one expects the peculiar velocity field to derive from a
potential too if it was so initially.
In Fourier space
it is then possible to get the very same system as Eqs. (\ref{FourierCont}-\ref{FourierEul})
but changing $\epsilon(t)$ to a $k$ dependent function that reads
\begin{equation}
\epsilon(k,t)=\frac{2\beta^2}{1+m^2a^2/k^2}
\end{equation}
which captures the whole effects of modified gravity. As a consequence  the linear growth rate is now $k$ dependent.

\subsection{The linear solution}

\begin{figure}[htbp]
\epsfig{file=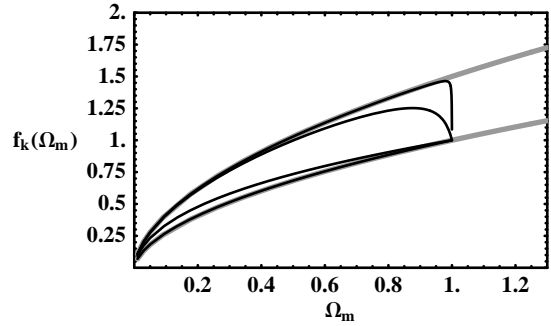,scale=0.6}
\caption{Dependence of the function $f_{k}$, defined in Eq. (\ref{fkdef}) as a function of $\Omega_{m}$ as given by Eq. (\ref{fksolution}). The solid grey lines show the behaviour of $f_{k}$
for standard gravity (lower curve) and modified gravity (upper curve) for $2\beta^2=1$  assuming the the wave mode
$k$ is such that $k\gg ma_{0}$.  The black lines show the actual behaviour of $f_{k}(\Omega)$  as a function of $\Omm(a)$ following its variation  during the expansion of the universe for $k/(ma_{0})=100$, $10$, $1$ and $0.1$ from the upper curve to the lower curve. Depending on $k$, the transition from the $\beta=0$ to the $\beta=1/\sqrt{2}$ solution
takes place at different time.
}
\label{fOm}
\end{figure}

When linearised this system leads to the following equation for the growth rate of the fluctuations,
\begin{equation}
\ddot D(k,t)+2H\dot D(k,t)-\frac{3}{2}H^2\Omm(1+\epsilon(k,t))D(k,t)=0\,.\label{lingrowth}
\end{equation}
Let us denote by $D_{+}$ the solution of this equation which grows with time. We can then define $\eta_{k}=\log D_{+}(k,t)$
and the factor $f_{k}(t)$ as
\begin{equation}
f_{k}=\frac{\dd \log D_{+}(k,t)}{\dd \log a(t)}.\label{fkdef}
\end{equation}
Thus we have, for any time dependent function $X$,
\begin{equation}
\dot X=f_{k} H X',
\end{equation}
where $X'$ is the derivative of $X$ with respect to $\eta_{k}$. We note that the linearized continuity equation reads
$\theta(k)=-f_{k}\delta(k)$ and the equation (\ref{lingrowth}) can be rewritten as,
\begin{equation}
2+f_{k}-\frac{3}{2}\frac{\Omm}{f_{k}}(1+\epsilon(k))+f_{k}'+\frac{\dot H}{H^2}=0.
\end{equation}
The resulting behaviour of $f_{k}$ is shown in Fig. \ref{fOm} assuming ${\dot H}/{H^2}$ is the one of a $\Lambda-$CDM
universe, i.e. ${\dot H}/{H^2}=-3/2\Omm$. In this case one has,
\begin{equation}
f_{k}\Omm'=3\Omm(\Omm-1).
\end{equation}
The evolution equations for $f_{k}$ and $\Omm$ then lead to,
\begin{equation}
\frac{\dd f_{k}}{\dd \Omm}=\frac{3/2\,\Omm(f+1+\epsilon)-f_{k}(2+f_{k})}{3\,\Omm(\Omm-1)}
\label{EqfOm}
\end{equation}
with the initial condition that $f_{k}$ is constant when $\Omm\to 1$. This equation can actually be solved explicitly for fixed values of $\epsilon$.
The solution reads,
\begin{eqnarray}
\hspace{-1cm}f_{k}(\Omm)&=&\frac{1}{4 \, _2F_1\left(\frac{f_+}{3},\frac{1}{3} \left(f_++2\right);\frac{1}{6} \left(4 f_++7\right);1-\frac{1}{\Omega_m}\right) \left(4 f_++7\right)
   \Omega_m}\nonumber\\
&&\hspace{-3cm}\times
   \left\{12 \, _2F_1\left[\frac{1}{3} \left(f_++3\right),\frac{1}{3} \left(f_++5\right);\frac{1}{6} \left(4 f_++13\right);1-\frac{1}{\Omega_m}\right] \left(\epsilon +f_++1\right)
   \left(\Omega_m-1\right)\right.\nonumber\\
&&\hspace{-3cm}\ \ \ \ +\left.   4 \, _2F_1\left(\frac{f_+}{3},\frac{1}{3} \left(f_++2\right);\frac{1}{6} \left(4 f_++7\right);1-\frac{1}{\Omega_m}\right) \left(6 \epsilon +5
   f_++6\right) \Omega_m\right\}
   \label{fksolution}
\end{eqnarray}
with
\begin{equation}
f_+=\frac{(25+24 \epsilon)^{1/2}-1}{4}
\end{equation}
corresponding to the Einstein-de Sitter value of $f$ in the presence of modified gravity.

This solution is illustrated in Fig. \ref{fOm}. The two grey lines correspond to constant values of $\epsilon$, either
$\epsilon=0$ (which corresponds to standard GR, lower line) or $\epsilon=1$ (small scale modified gravity with $2\beta^2=1$, upper line). It is found that the general solution, for a constant value of $\epsilon$ is  very close to~\footnote{This form can be obtained
assuming $f_{k}\approx f_{+}\,\Omega^{\nu}$ and expanding Eq. (\ref{EqfOm}) to first order in $\omega$ with $\Omega=1+\omega$ which gives the value of $\nu$.},
\begin{equation}
f(\Omega_m)\approx f_+\ \Omega_m^{\frac{2(2+f_+)}{4 f_++7}}.\label{fkexp}
\end{equation}
When $k$ is comparable to or larger than  $k_{c}=m\,a_{0}$ ($a_{0}$ is the current value of the expansion factor) the function
$f_{k}$ switches for one curve to the other as shown in Fig \ref{fOm} with the solid black curves. It is to be noted that this
result does not follow the form of Eq. \ref{fgamform}, a feature already noticed in~\cite{Brax:2009ab}.

\subsection{The mode coupling evolution}

Similarly to the first section, let us define the reduced velocity mode by dividing it by $f_{k}(t)$ thus introducing
$\thetar(k)=\theta(k)/f_{k}$.
The continuity and Euler equations now become
\begin{eqnarray}
\delta'(k)+{\thetar}(k)=-\Dirac(\vk-\vk_{1}-\vk_{2})\alpha(\vk_{1},\vk_{2})\,\delta(k_{1})\,\thetar(k_{2})\frac{f_{k_{2}}}{f_{k}}\label{ContExp}\\
\thetar'(k)-\left(1-\frac{3}{2}\frac{\Omm}{f_{k}^2}(1+\epsilon(k))\right)\thetar(k)+\frac{3}{2}\frac{\Omm}{f_{k}^2}(1+\epsilon(k))\delta(k)=\nonumber\\
\hspace{3cm}-\Dirac(\vk-\vk_{1}-\vk_{2})\beta(\vk_{1},\vk_{2})\thetar(k_{1})\thetar(k_{2})\frac{f_{k_{1}}f_{k_{2}}}{f_{k}^2}\label{EulerExp}
\end{eqnarray}
The evolution of the nonlinear couplings will then depend on ${\Omm}/{f_{k}^2}\,(1+\epsilon(k))$.
In general this system can be solved explicitly in a regime where all the $f_{k}$'s  are independent on $k$ and where ${\Omm}/{f_{k}^2}\,(1+\epsilon(k))$ is independent on both $k$ and time.

\begin{figure}[htbp]
\epsfig{file=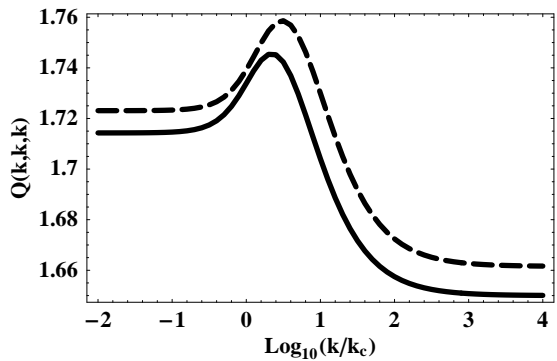,scale=0.6}
\caption{Amplitude of the reduced bispectrum of the density field  for equilateral configurations as a function of $k/k_{c}$ for an Einstein-de Sitter background (dashed line) and a $\Lambda-$CDM background (solid line). For low values of $k$, the result expected for standard GR is recovered. For high values of $k$, modified gravity effects take place. }
\label{Equi}
\end{figure}

In the standard GR regime and when the universe is still dominated  by its matter content we have $\Omm/f^2\approx 1$.
This is the regime we have at early time, that is at large enough redshift. In the context of standard cosmology, it happens that, for a $\Lambda$-CDM background, the solution stays close to $f=\Omm^{0.5}$  so that $F_{2}$ and all higher order functions $F_{p}$ are only weakly evolving with time. They remain close to the Einstein--de Sitter solution.
In the context of modified gravity regime this will still be the case for large enough scales  (i.e.  $k\ll k_{c}$) since then
$\epsilon(k,t)$ is small (and has always been so assuming that $m\,a$ is decreasing with time).
As a result the higher order terms, to all orders, are expected to remain formally
the same (as for the Einstein-de Sitter case) when they are expressed in terms of the  linear solution.
This will be illustrated on the left hand plateau of Fig. \ref{Equi}.

Another regime of interest is when all scales  are well within the modified gravity regime. In this case $\epsilon(k,t)$
is both independent of $k$ and of time (it is simply $2\beta$) and, according to   Eq. (\ref{fkexp}),
${\Omm}/{f_{k}^2}\,(1+\epsilon(k))$ is also found to be roughly constant but with a slightly different value. As a consequence the formal expression of $F_{2}$ show in Eq. (\ref{PrF2Exp}) is still valid provided $\nu_{2}$ is adequately calculated. In the $\Omm\to 1$ limit we have
\begin{equation}
\nu_{2}(\epsilon)=\frac{2(8+9\xi)}{3(4+3\xi)}\ \ {\rm with}\ \ \xi=\frac{1+\epsilon}{f_{+}^2},\label{nu2exp}
\end{equation}
which incidentally varies from $34/21$ to $14/9$ when $\beta$ varies from $0$ to $\infty$. Quite surprisingly $\nu_{2}$ remains finite even when the effect of self-gravity is made arbitrarily large.
This induces a change, although modest, for the tree-order bispectrum as we will see later.

In general however one is interested in mode coupling effects for modes that may not be in the same regimes and for which
we do not have necessarily $f_{k}=f_{k_{1}}=f_{k_{2}}$. The general form $F_{2}$ is then changed and a case by case numerical integration is then necessary (see \cite{2004astro.ph..9224B}).

\subsection{Bispectra}

The effect of modified gravity is illustrated in Fig. \ref{Equi} which gives the expression of the reduced
bispectrum in case of equilateral configurations (e.g. $k_{1}=k_{2}=k_{3}$). Note that in this case, the result is independent of
the shape of the power spectrum. We can see the effect of modified gravity compared  to the effects of the background evolution (dashed to solid line). For the dashed line the asymptotic plateaus  can be obtained from the form (\ref{PrF2Exp}) and the values
of $\nu_{2}$ given by Eq. (\ref{nu2exp}) for $\epsilon=0$ and $\epsilon=1$.

\begin{figure}[htbp]
\epsfig{file=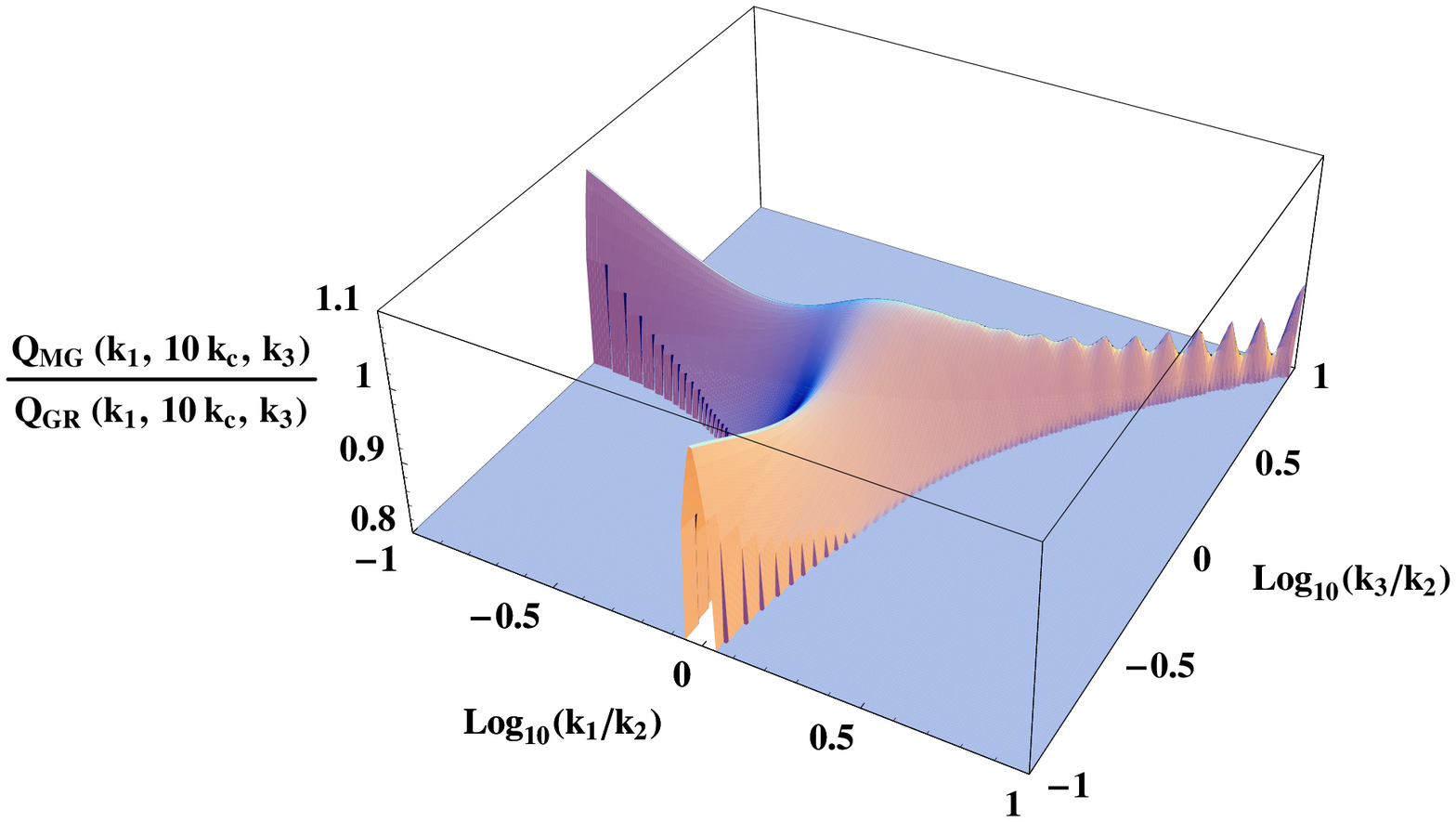,scale=.35}\hspace{.1cm}\epsfig{file=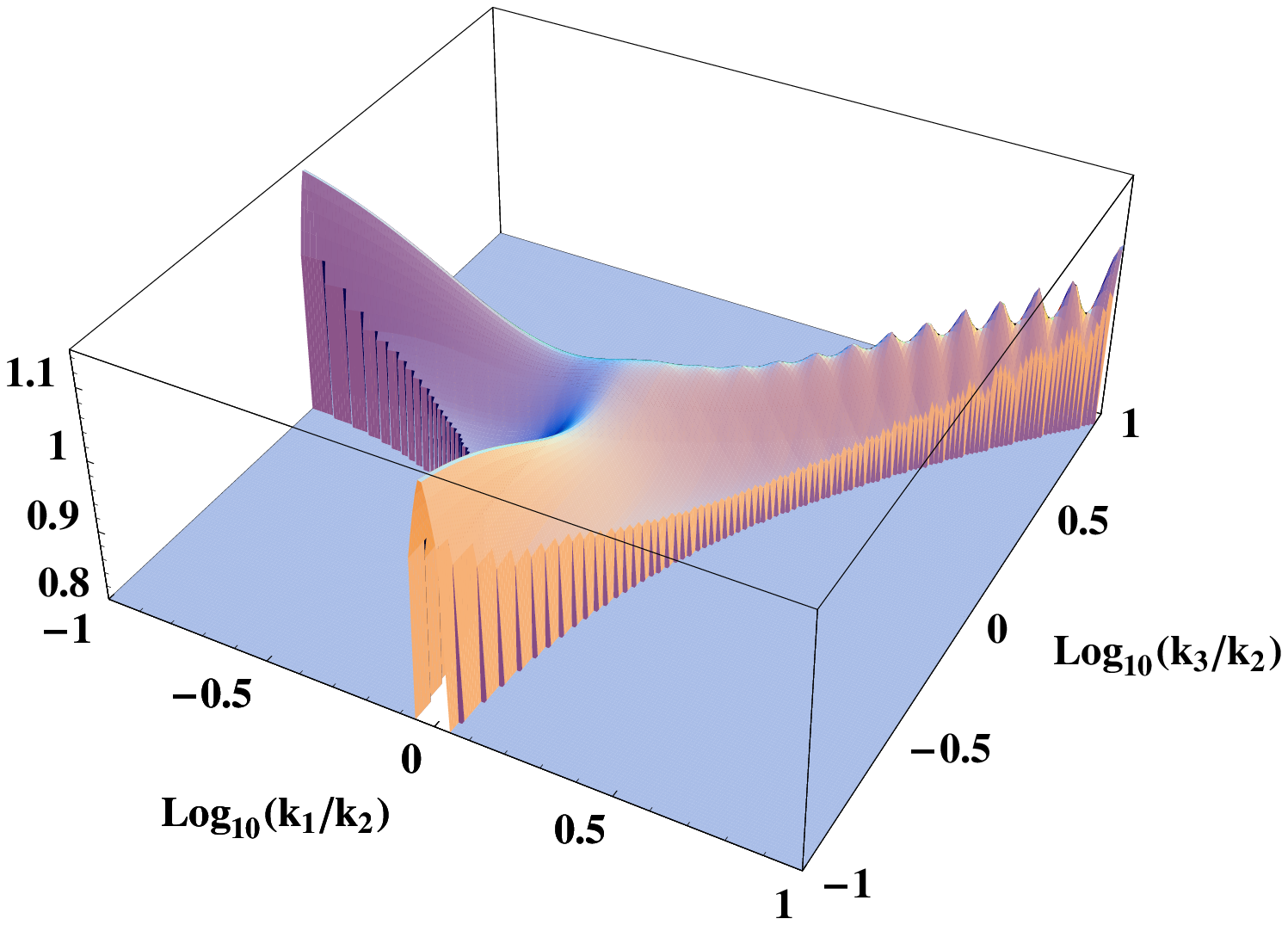,scale=.35}
\caption{Amplitude of the reduced bispectrum $Q(k_{1},k_{2},k_{3})$ of the density field
for the modified gravity divided by the expected result for
standard gravity as a function of $k_{1}/k_{2}$ and $k_{3}/k_{2}$ for $k_{2}=10 k_{c}$. The left panel corresponds to $P(k)\sim k^{-1}$ and the right panel to $P(k)\sim k^{-2}$. Note that the irregular patterns on the right sides of the figures are due to inadequate interpolation effects.}
\label{BiSpec}
\end{figure}

Finally in Fig. \ref{BiSpec} a more general $k$ dependence of the bispectrum is shown and compared to the standard GR results. It shows that for some elongated configurations modified gravity can depart from standard GR by as much as $15\%$. Notice that the effect is expected to be much larger for the statistical properties of $\theta$ as we will see it in a more complete analysis. The computations have been done for two different hypotheses for the power spectrum shape, $P(k)\sim k^{-1}$ (left panel) and $P(k)\sim k^{-2}$ (right panel). The two panels exhibit subtle differences that are due the fact that the different contribution to the bispectrum are
weighted slightly differently for non-equilateral configurations.

\section{Modified Gravity with a General Scalar-Tensor Theory}
\label{NonLinearEffects}
\newcommand{\deltam}{\delta_{\rm m}}

\subsection{Modified gravity}

For a general scalar-tensor theory with a field dependent  coupling $\beta$, the perturbation equations are slightly modified from the ones used in the linear case. First of all the mass is sensitive to the perturbations and therefore the third derivative of the potential plays a role. Similarly, the coupling can be perturbed and thence its derivative is also relevant.
First of all in the Einstein frame the Einstein equations read
\begin{equation}
G_{\mu\nu}\equiv R_{\mu\nu}-\frac{R}{2} g_{\mu\nu}= \kappa_4^2 (T_{\mu\nu}^m + T^{\varphi}_{\mu\nu})
\end{equation}
where $T^m_{\mu\nu}$ is the matter energy momentum and $T^{\varphi}_{\mu\nu}$ is the scalar field energy momentum tensor
\begin{equation}
T^{\varphi}_{\mu\nu}=2\partial_\mu\varphi\partial_\nu\varphi -(V(\varphi) + (\partial \varphi)^2) g_{\mu\nu}
\end{equation}
where we are working with the normalised field $\varphi$ for simplicity. The Klein-Gordon equation is simply
\begin{equation}
\nabla^\mu\nabla_\nu \varphi= \frac{1}{2}( \partial_\varphi V -\beta_{\rm eff} (\varphi)T^m)
\end{equation}
where $T^m$ is the trace of the energy momentum tensor and $\beta_{\rm eff}={\beta (\phi)}/{k(\phi)}$
Using the Bianchi identity $\nabla _\mu G^{\mu\nu}=0$ we find the non-conservation equation
\begin{equation}
\nabla_\mu T^{\mu\nu}_m = \beta_{\rm eff} (\varphi) T^m \partial^\nu \varphi\,.
\end{equation}
In the case of a pressure-less fluid with
\begin{equation}
T^{\mu\nu}_m=\rho^E u^\mu u^\nu
\end{equation}
where $u^\mu={dx^\mu}/{d\tau}$ and $\tau$ is the proper time along the  evolution of CDM particles with $u^2=-1$, we find that
\begin{equation}
\dot \rho^E +3 h \rho^E= \rho^E \beta_{\rm eff} (\varphi) \dot \varphi
\end{equation}
where $\dot F= u^\mu \nabla_\mu F$ for any tensor and the local Hubble rate is $3h=\nabla_\mu u^\mu$. Similarly the Euler equation becomes
\begin{equation}
\dot u^\mu +\beta(\varphi) \dot \varphi u^\mu = -\beta_{\rm eff} (\varphi) \partial^\mu \varphi\,.
\end{equation}
In the following we will neglect the time variation of the scalar field compared to the spatial gradients. Hence in the non-relativistic limit the Euler equation becomes,
\begin{equation}
\dot u^i = -\beta_{\rm eff}(\varphi) \partial^i \varphi,
\end{equation}
implying that the CDM particles feel the presence of a scalar force of coupling strength $\beta_{\rm eff} (\varphi)$. It will be  very useful to redefine the matter density
\begin{equation}
\rho^E= A(\varphi) \rho
\end{equation}
so that the conservation of matter equation is satisfied (notice that a dot here stands for the derivative with respect to proper time)
\begin{equation}
\dot \rho +3 h \rho=0
\end{equation}
or equivalently
\begin{equation}
\nabla_\mu (\rho u^\mu)=0.
\end{equation}
This is the usual conservation of matter in cosmology. The conserved matter density is not $\rho^E$ but a rescaled density depending on the Weyl factor $A(\varphi)$.

Finally the Klein-Gordon equation can be written as
\begin{equation}
\nabla^\mu \nabla_\mu \varphi= 4\pi G_N \partial_\varphi V_{\rm eff}(\varphi),
\end{equation}
where the effective potential is simply
\begin{equation}
V_{\rm eff}(\varphi)= V(\phi) + A(\phi) \rho
\end{equation}
and the conserved density $\rho$ is $\varphi$-independent.
This is the expression we have used already for chameleon and dilaton models.
These equations lead to the proper set of perturbation equations in the general case when complemented with the Poisson equation with the Newtonian
potential depending on $A(\varphi)\rho$.

In conclusion, we have found that the Euler equation is modified by the presence of a scalar force while conservation of matter is still valid. The conserved matter density must be dressed with the conformal factor $A(\varphi)$ to act as a source for the Newton potential in the Poisson equation.

\subsection{Perturbation equations}

The Euler equation can be written explicitly in the single flow approximation, neglecting pressure terms:
\begin{eqnarray}
\frac{\partial \vu_{i}(\vx,t)}{\partial t}+H(t) \vu_{i}(\vx,t)+\frac{1}{a(t)}\vu_{j}(\vx,t)\vu_{i,j}(\vx,t)=\nonumber\\
\hspace{2cm}-\frac{1}{a(t)}\left(\Phi(\vx,t)_{,i}+\beta_{\rm eff}(\bar \phi+\delta \phi)\varphi(\vx,t)_{,i}\right)
\end{eqnarray}
where $\Phi$ is the usual Newton potential (the one that enters in the metric) and $\phi(\vx,t)$ the scalar field
that couples to the matter field. Both fields are coupled to the matter fluctuations.
For the Newton potential $\Phi$, we have the usual Poisson equation,
\begin{equation}
\frac{1}{a^2(t)}\Delta\Phi(\vx,t)=4\pi GA(\bar\phi) \rhob(t)\,\deltam(\vx,t)
\end{equation}
where
\begin{equation}
\deltam(\vx,t)= \delta(A(\phi) \rho)
\end{equation}
which depends on  both the scalar and matter perturbations.
Using the Klein-Gordon equation
\begin{eqnarray}
\frac{1}{a^2(t)}\Delta \varphi(\vx,t)+4\pi G  \left[V_{\rm {eff},\varphi}(\bar \phi)-V_{\rm {eff},\varphi}(\phi)\right]=\nonumber\\
\hspace{5cm}4\pi G\ \beta_{\rm {eff}}(\phi) A(\bar\phi)  \rhob(t)\,\deltam(\vx,t).
\end{eqnarray}
where we have defined
\begin{equation}
\beta_{\rm eff}(\phi)= \frac{\beta(\phi)}{k(\phi)}
\end{equation}
 we find that the Klein-Gordon equation, the Poisson equation and the Euler equation
involve $\deltam$ and not $\delta\rho/\rho$, i.e.  the Einstein frame energy density. As a result the perturbation variables can be conveniently chosen to be $\delta\varphi$, $\deltam$ and $\theta$. Of course this is not true of the matter conservation equation as the Einstein frame energy density is not conserved in general. On the other hand here the Klein-Gordon equation leads to the estimate for the $p-$th order in perturbation theory
\begin{equation}
\delta\varphi^{(p)}={\cal O}( \frac{a^2H^2}{k^2 +a^2m^2}(\deltam^{(1)})^p)
\end{equation}
implying that the discrepancy between $ \delta\rho/\rho$ and $\deltam$ is negligible when $m\gg H$. This condition is satisfied for the models we consider as the background mimics $\Lambda$-CDM and the background value of the scalar field is an attractor. In the following, we will use the perturbed equations as a function of $\deltam$.

\subsection{Second order perturbations}

\begin{figure}[htbp]
\epsfig{file=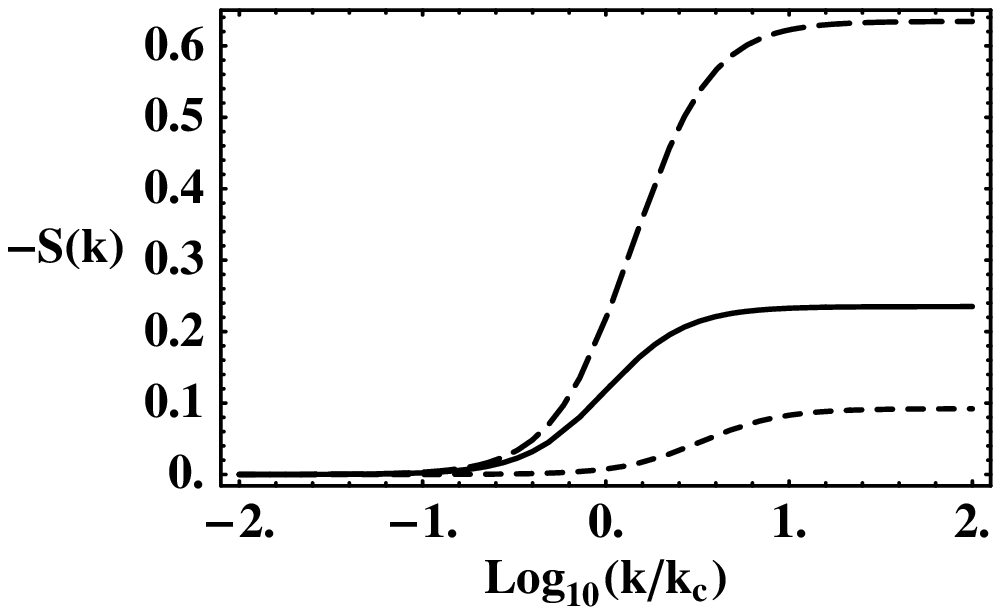,width=4cm}\hspace{.1cm}
\epsfig{file=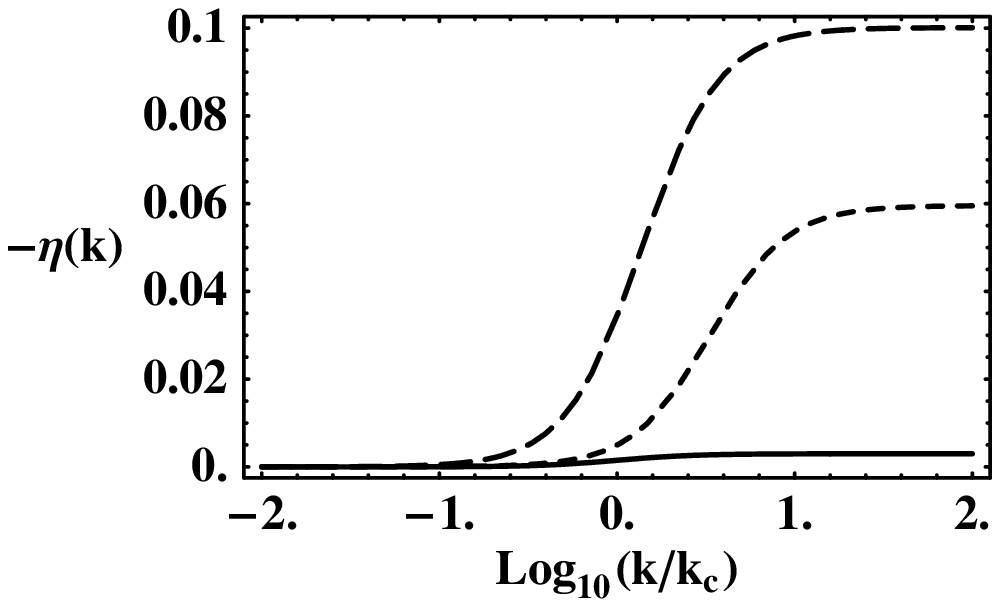,width=4cm}\hspace{.1cm}
\epsfig{file=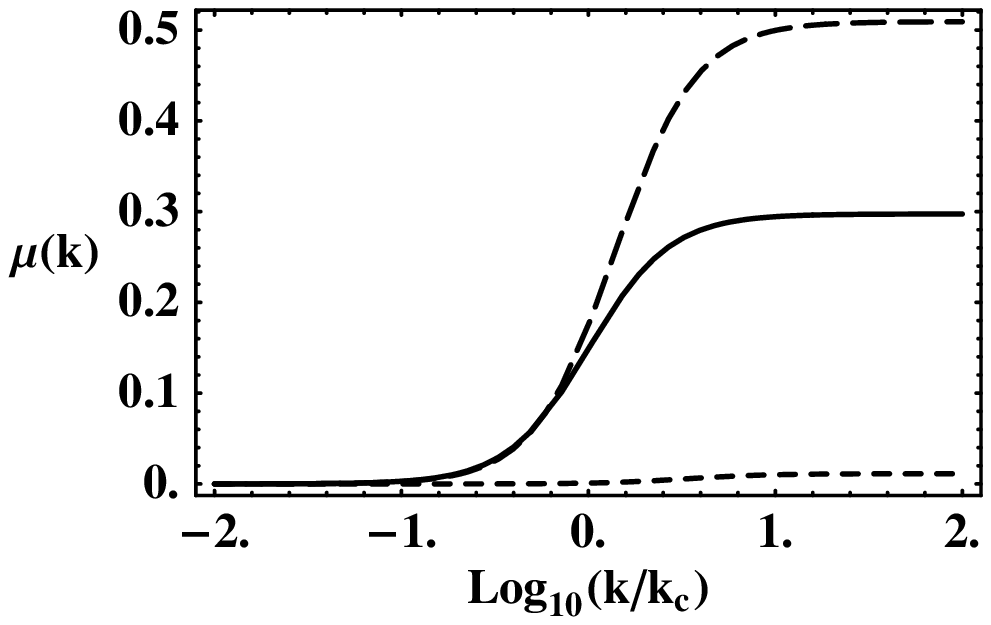,width=4cm}
\caption{Dependence on $k$ of the parameters $S(k)$, $\eta(k)$ and $\mu(k)$ for $\eta_\Omega=0$ (solid lines), $\eta_\Omega=-1$ (long dashed) and $\eta_\Omega=-2$ (short dashed).}
\label{ParamEvol}
\end{figure}

We now focus our attention on the second order perturbation equations.
The first step is to take the the divergence of the Euler equation:
\begin{equation}
\hspace{-2cm}
(\dot{aH \theta}) + aH^2 \theta +\frac{1}{a} \nabla_i(u_j \nabla_i u_j) = -\frac{1}{a}(\Delta \Phi +\beta_{\rm eff}(\phi) \Delta \delta\varphi+ \nabla_i\beta_{\rm eff}(\phi)\nabla_i\delta \varphi).
\end{equation}
Working to second order only, this equation simplifies and becomes
\begin{eqnarray}
\frac{d}{d t}({aH \theta}) + aH^2 \theta +\frac{1}{a} \nabla_i(u_j \nabla_i u_j) +\frac{1}{a}(\Delta \Phi + \beta_{\rm eff}(\bar \phi) \Delta \delta\varphi)\nonumber\\
\hspace{4cm}=-\frac{1}{a}(\gamma_{\rm eff}(\bar \phi)\nabla_i(\delta\varphi \nabla_i\delta \varphi))
\end{eqnarray}
where
\begin{equation}
 \ \gamma_{\rm eff}(\bar\phi)=\frac{ \beta_{\rm eff,\phi}(\bar\phi)}{k(\bar \phi)}\,.
\end{equation}
Let us now analyse the perturbed Klein-Gordon equation at the second order:
\begin{eqnarray}
(-\frac{\Delta}{a^2}+ m^2 (\bar \phi))\delta \varphi+\frac{3}{2}\Omega_m H^2 \beta_{\rm eff}(\bar\phi)A(\bar\phi) \deltam=\nonumber\\
\hspace{3cm}-\frac{3}{2}H^2 \Omega_m A(\bar\phi)\gamma_{\rm eff}(\bar \phi) \delta \varphi \deltam - u(\bar\phi)\frac{\delta\varphi^2}{2}
\end{eqnarray}
where
\begin{equation}
u(\bar\phi)=4\pi G\frac{d^3V_{\rm eff}}{d\varphi^3}.
\end{equation}
We then get the Fourier components of the first and second order perturbations of the field.
At first order we have,
\begin{equation}
\delta \varphi^{(1)}= \frac{a^2 H^2}{k^{2}}S(k)\deltam^{(1)}
\end{equation}
with the function $S(k)$ defined as
\begin{equation}
S(k)= -\frac {3}{2}\Omega_m \beta_{\rm eff}(\bar \phi) \frac{1}{1+a^2m^2(\bar \phi)/k^2}.\end{equation}
The behaviour of this function is depicted in Fig. \ref{ParamEvol}.
At second order the Klein-Gordon equation reads
\begin{eqnarray}
\frac{k^2}{a^2}\delta \varphi^{(2)}&=& H^2\deltam^{(2)}
 -\frac {3}{2}\Omega_m H^2 A(\bar\phi) \gamma_{\rm eff}(\bar \phi)\frac{\delta\varphi^{(1)} \deltam^{(1)}}{1+a^2m^2(\bar \phi)/k^2}
 \nonumber \\
&& -\frac{u (\bar \phi)}{2}\frac{(\delta\varphi^{(1)})^2}{1+a^2m^2(\bar \phi)/k^2}
\end{eqnarray}
where we have used the notation for the product of perturbations
\begin{equation}
AB=\Dirac(\vk-\vk_1-\vk_2) A(k_1) B(k_2).
\end{equation}

\begin{figure}[htbp]
\epsfig{file=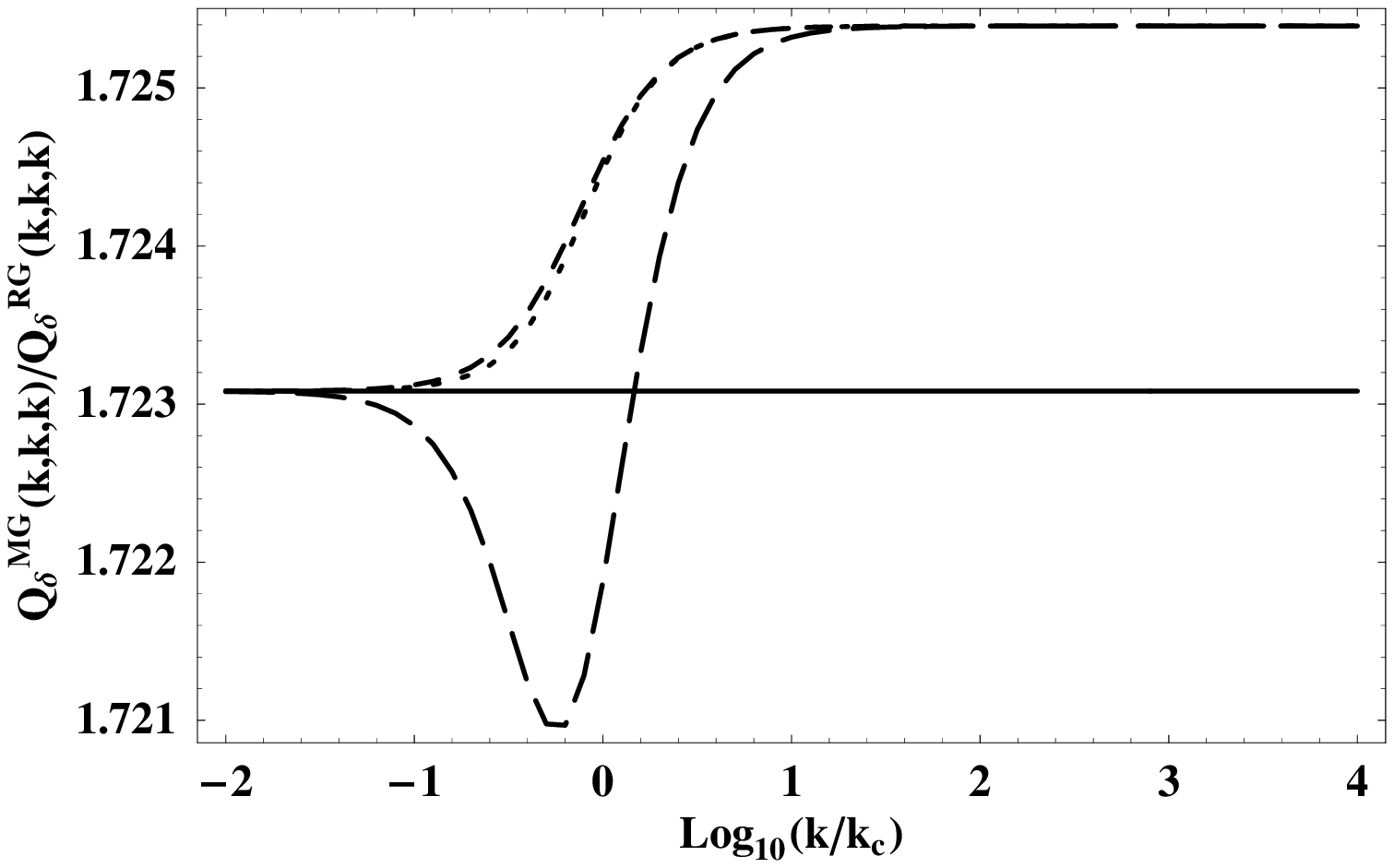,width=4.cm}\hspace{.1cm}
\epsfig{file=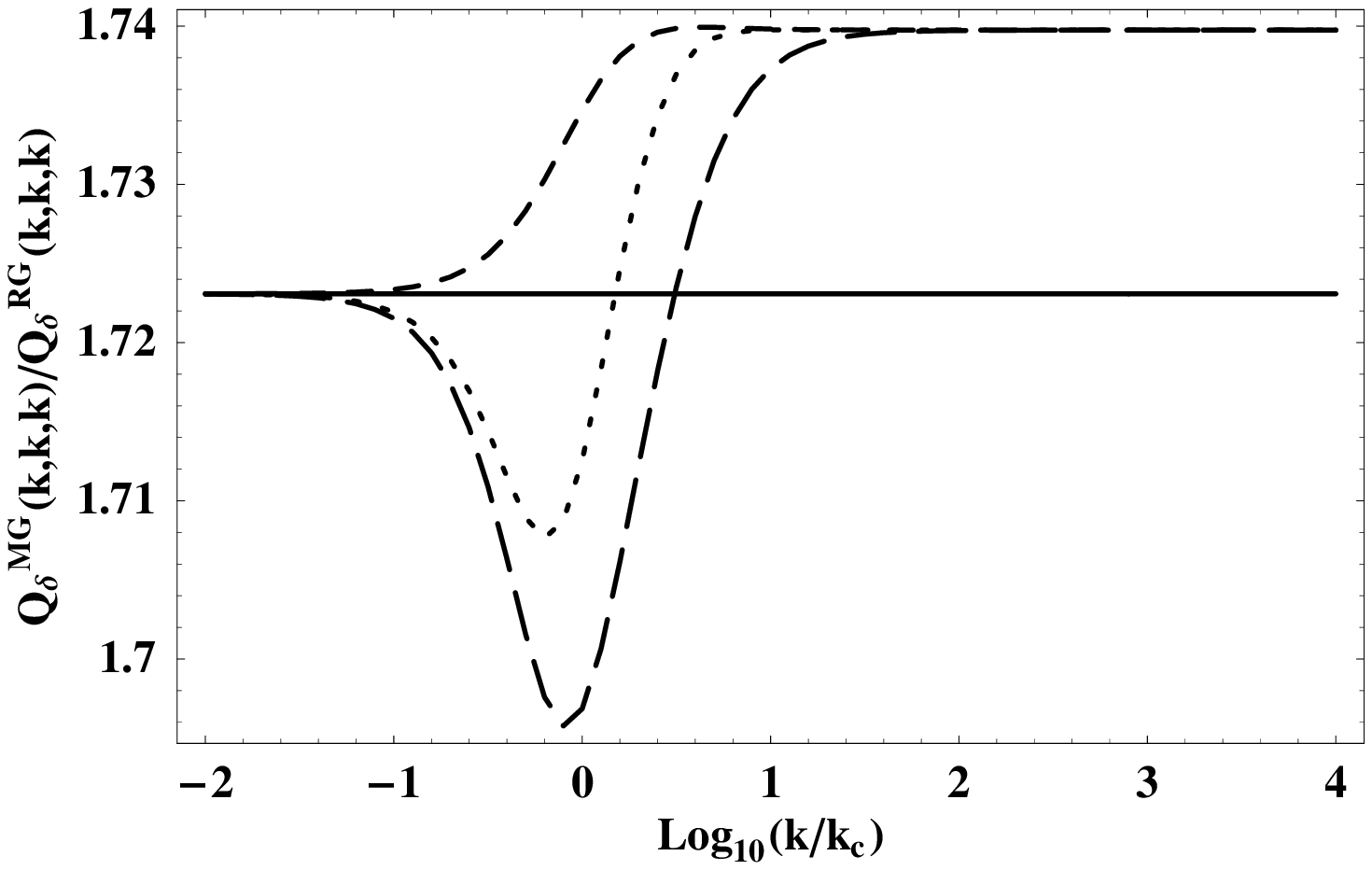,width=4.cm}\hspace{.1cm}
\epsfig{file=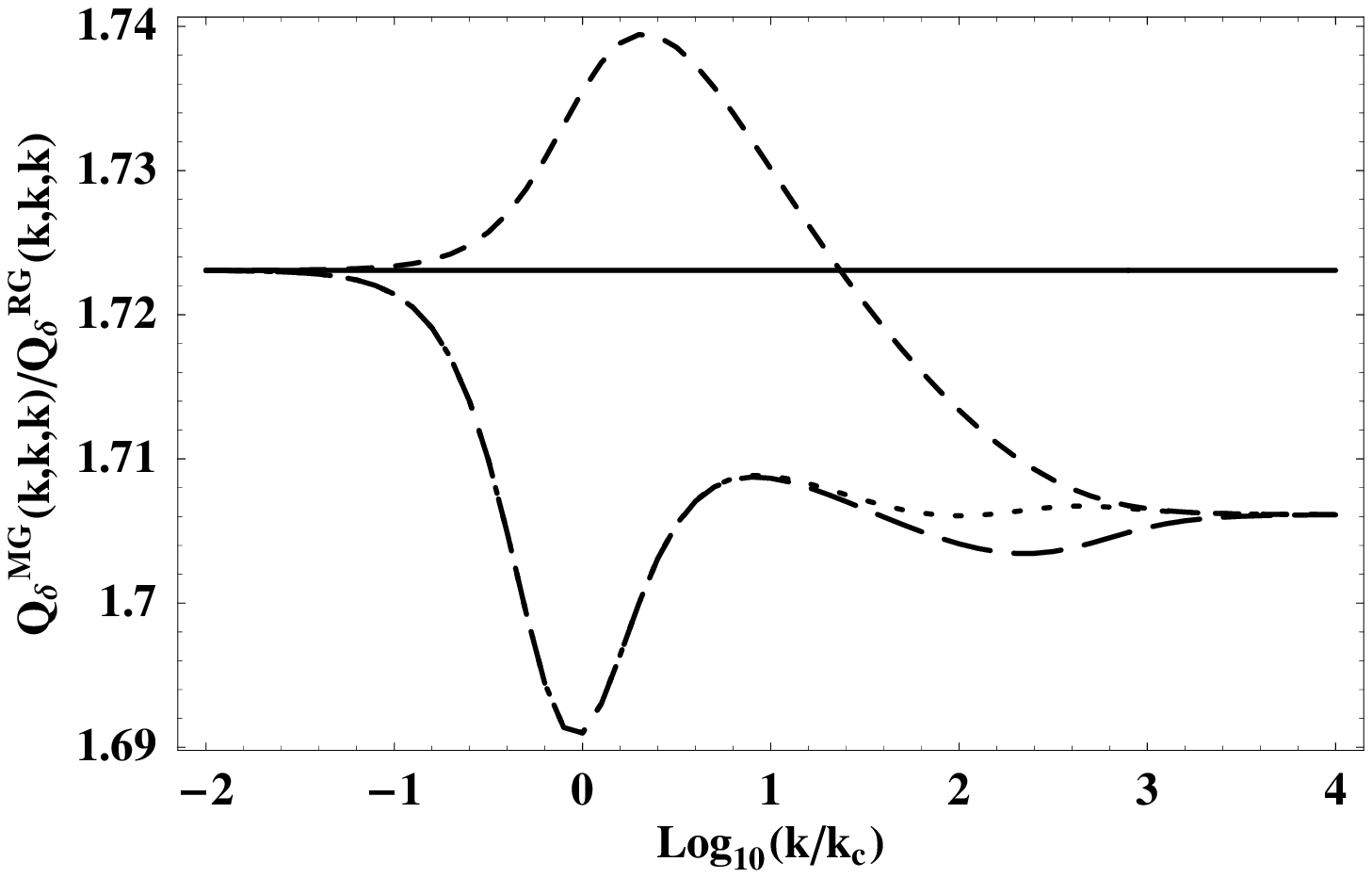,width=4.cm}\\
\epsfig{file=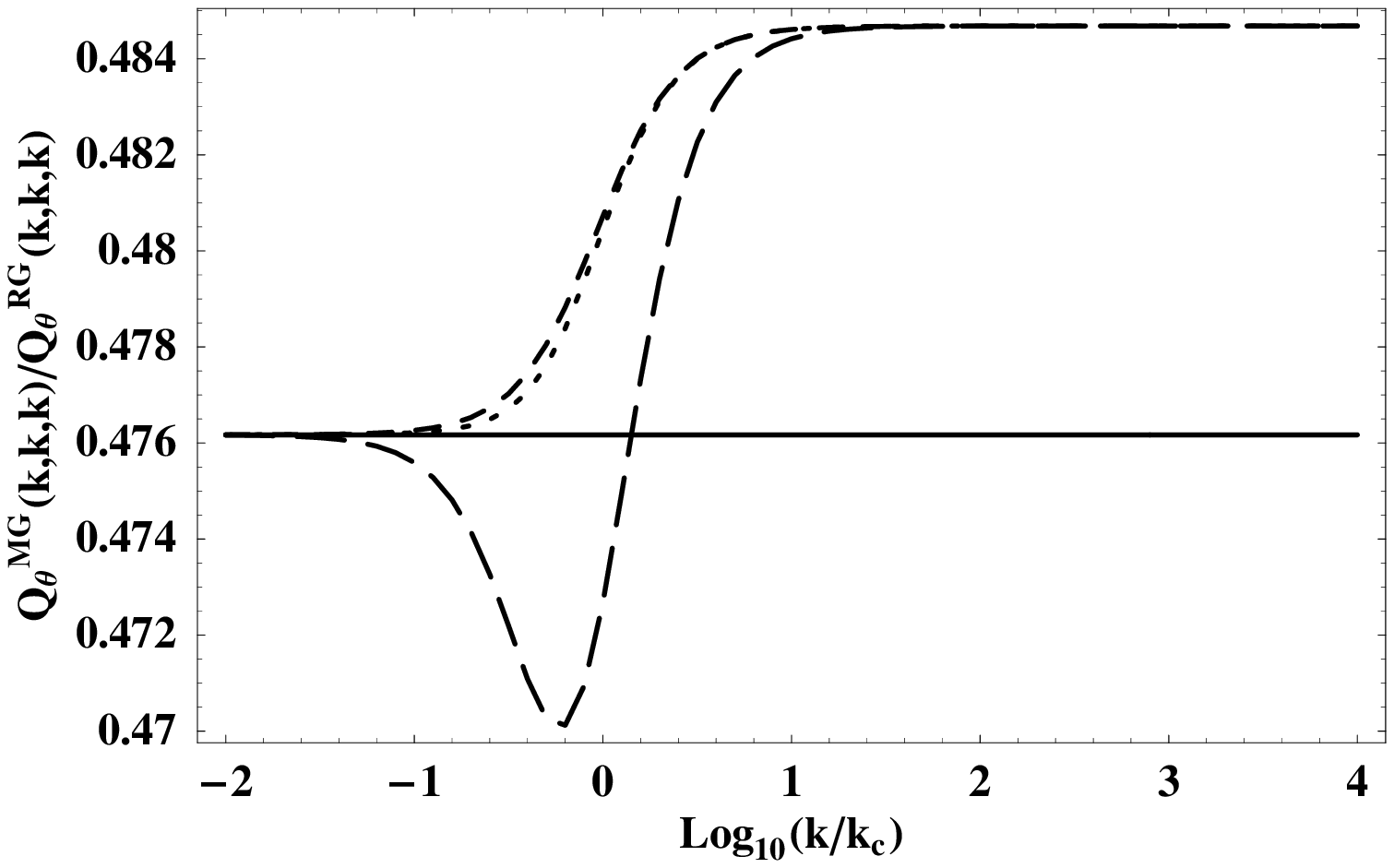,width=4.cm}\hspace{.1cm}
\epsfig{file=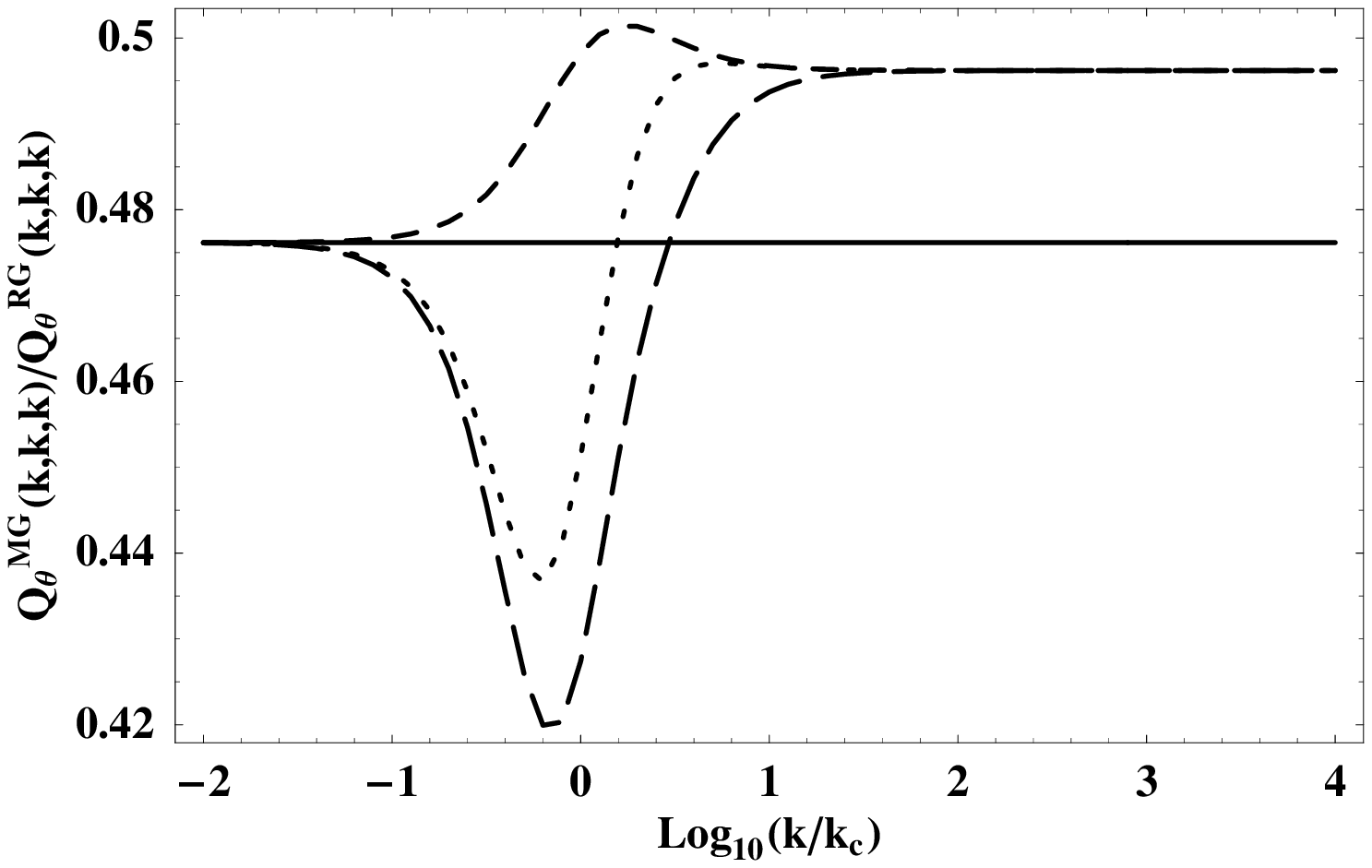,width=4.cm}\hspace{.1cm}
\epsfig{file=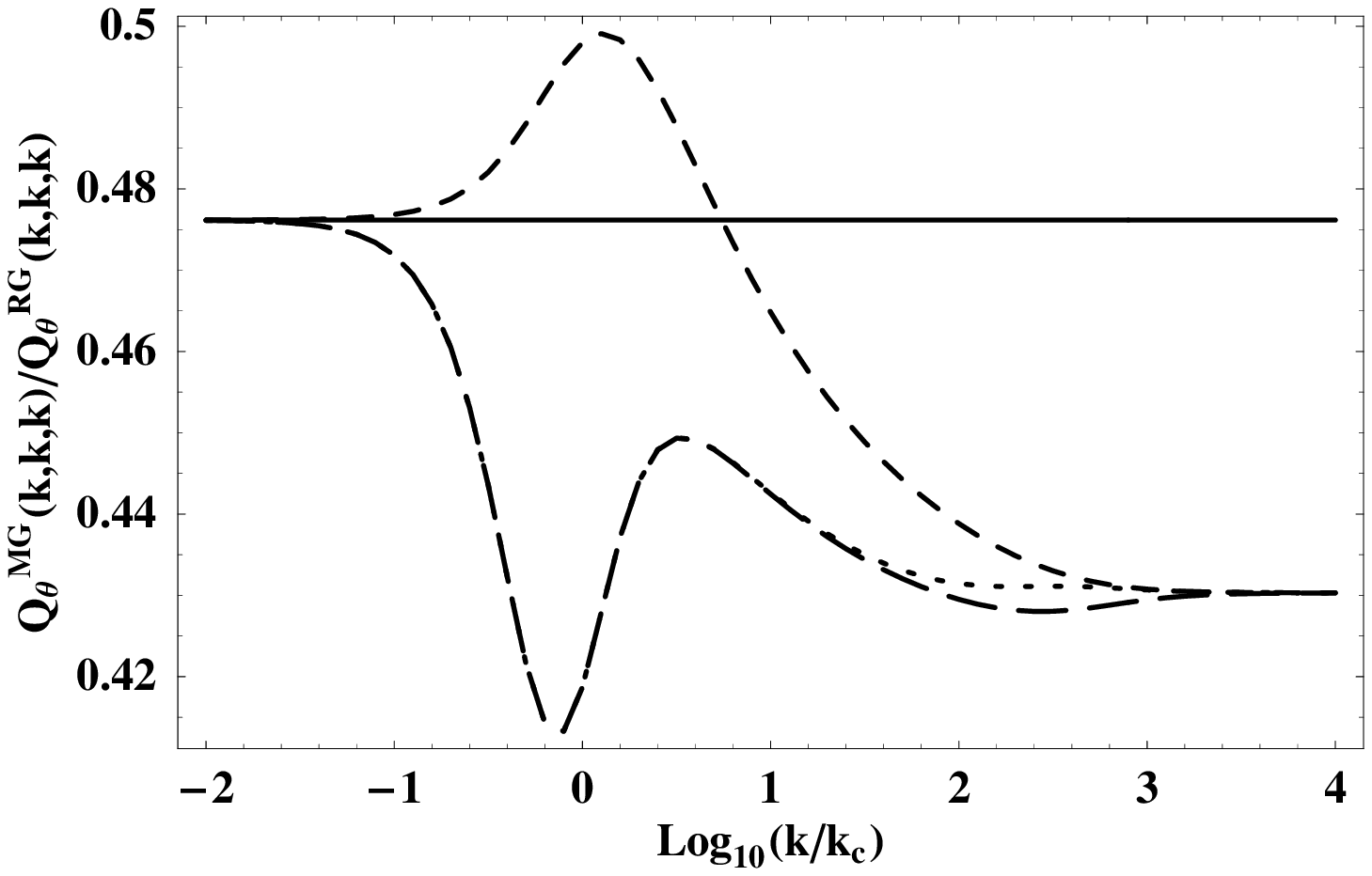,width=4.cm}
\caption{The density (top row) and reduced velocity divergence (bottom row) bispectra for equilateral configurations as a function of scales for $\lambda=1$ left panels, $\lambda=10$, middle panels and $\lambda=1000$, right panels. The solid line is the
General Relativity prediction.  The modified gravity model used here corresponds to the long dashed lines. The short dashed line is obtained when the extra couplings that appear in the Euler equation are dropped. They reproduce the large $k$ behavior. The dotted line correspond to the case when only the intrinsic coupling of the $\phi$ field is preserved. It gives the dominant contribution when $\lambda$ is large (right panels) but a negligible one
when $\lambda$ is small (left panels).
}
\label{Bispectra}
\end{figure}

We are now in position to write the Euler equation at second order. It takes the form:
\begin{eqnarray}
\frac{1}{H}\dot\theta^{(2)}+(2+\frac{\dot H}{H^2})\theta^{(2)} +\frac{3}{2}\Omega_m (1+\epsilon(k)) \deltam^{(2)}&=& -\beta(\vk_1,\vk_2) \theta^2 \nonumber\\
&&\hspace{-4cm}-\left[\mS_{\rm Eul.}(\vk_1,\vk_2) +\mS_{\rm Intr.}(\vk_1,\vk_2) \right](\deltam^{(1)})^2\label{EulerMod}
\end{eqnarray}
where we have introduced
\begin{equation}
\mS_{\rm Eul.}(\vk_1,\vk_2) =\frac{(\vk_2.\vk)}{k_{1}^2}\frac{a^2 m^2(\bar\phi)}{k_{2}^2}S(k_1)\eta(k_2)
\end{equation}
and
\begin{eqnarray}
\mS_{\rm Intr.}(\vk_1,\vk_2) &=&
\frac{a^2 m^2(\bar\phi)}{k_{2}^2}S(k)\eta(k_{2})\nonumber\\
&&+\frac{a^2 m^2(\bar\phi)}{k_{1}^2}\frac{a^2 m^2(\bar\phi)}{k_{2}^2} S(k_{1})S(k_{2})\mu(k)
\end{eqnarray}
with
\begin{equation}
\eta(k)=S(k)\frac{H^2 A(\bar\phi)\gamma_{\rm eff}(\bar\phi)}{m^2(\bar\phi)},\
\end{equation}
and
\begin{equation}
\mu(k)=S(k)\frac{H^2\,u(\bar \phi)}{m^4(\bar\phi)}\frac{1}{3\Omm}\,.
\end{equation}
The parameters $S(k)$, $\eta(k)$ and $\mu(k)$ have been chosen so that they remain finite when $k\approx k_{c}$.
We can now compare the Euler equation derived here with the ones obtained in GR. There are two main effects which signal the modification of gravity. The first one is the appearance of the $\epsilon$ term which was already captured in the $\gamma$-model and the linear approximation. In the context of scalar-tensor theories, we find that there are new contributing terms which are intrinsically due to the non-linearity of the theory.  Notice that at the second order level, the new terms in the perturbation equations amounts to a modification of $\beta(\vk_1,\vk_2)$:
\begin{equation}
\beta_{\rm eff}(\vk_1,\vk_2)= \beta(\vk_1,\vk_2)+\mS_{\rm Eul.}(\vk_1,\vk_2)+\mS_{\rm Intr.}(\vk_1,\vk_2)
\end{equation}
These terms originating  from the new $(\deltam^{(1)})^2$ contributions lead to new effects which will be analysed for dilaton models in the following section.

\subsection{Application to dilaton models}

The dilaton models are good benchmarks to test the influence of modified gravity at second order in perturbation theory. To do so, the mass and couplings must be known in the cosmological background, it turns out that the mass of the dilaton satisfies\cite{Brax:2010gi}
\begin{equation}
\frac{m^2 (\bar\phi)}{H^2}\approx \frac{3A_2}{2}\frac{\Omega_m +4\Omega_\Lambda}{(\lambda^{-2} +3(\frac{\Omega_m}{\Omega_\Lambda}+4)^{-2})}
\end{equation}
while
\begin{equation}
\beta(\bar\phi)\approx \frac{\Omega_\Lambda}{\Omega_m+4\Omega_\Lambda}\,.
\end{equation}
This allows one to deduce the other functions which appear in the perturbation equations.
Assuming that both $A_{2}$ and $\lambda^2$ are large, we have
\begin{eqnarray}
S(k)&\sim&-\frac{\sqrt{3}\Omm}{2},\ \ \hbox{when}\ k/k_{c}\to\infty\\
S(k)&\sim&-\frac{\sqrt{3} ({\Omm}-1)^2 {\Omm}}{(4-3 {\Omm})^3}\frac{k^2}{{A_{2}}},\ \ \hbox{when}\ k/k_{c}\to 0\,.
\end{eqnarray}
They are shown in the various panels of Fig. \ref{ParamEvol} for $\lambda=10$. It is to be noted that $S(k)$ is finite in the modified gravity domain. It takes the asymptotic
value of about 0.2 for $\eta_\Omega=0$.

The results of the second order modification of gravity are to a large extent determined by the asymptotic behaviour of the functions $\eta(k)$ and $\mu(k)$. Two cases
can be distinguished. When $\lambda$ is large the $\beta_{\rm eff}$ parameter becomes $\phi$ independent and $\gamma_{\rm eff}$  vanishes. In this case we have
\begin{eqnarray}
\eta(k)&\approx&\frac{2}{9\lambda^2}\frac{4-3\Omm}{(1-\Omm)^2}S(k)\\
\mu(k)&\approx&\frac{-2\sqrt{3}}{3}\frac{1}{\Omm(4-3\Omm)}S(k).
\end{eqnarray}
On the other hand, when $\lambda$ is small all parameters tend to be small and follow a hierarchical behaviour
leaving the terms containing  $\eta$ as the main coupling effect. In this case we have
\begin{eqnarray}
S(k)&\approx&-\lambda\frac{3\Omm(1-\Omm)}{2(4-3\Omm)}\frac{1}{1+a^2m^2(\bar\phi)/k^2}\\
\eta(k)&\approx&\frac{2}{3}\frac{1}{4-3\Omm}S(k)\\
\mu(k)&\approx&-2\frac{1-\Omega_m}{\Omm(4-3\Omm)^2}S(k).
\end{eqnarray}
All the extra couplings actually vanish in the modified gravity regime ($k\gg k_{c}$) because
of the $a^2m^2/k^2$ factors that enter in the expression of the coupling functions.

It is also worth noting that the extra coupling functions $\mS_{\rm Intr.}$ and $\mS_{\rm Eul.}$ all vanish when $\vk_{1}+\vk_{2}=0$.
In these cases the modifications of gravity reduce to the ones of the linear theory. As a result the effect of the non-linearity of modified gravity are only present in an intermediate regime around $k_c$.


The resulting shape of the bispectra are presented in details in the following (in particular in the appendix). The benchmark model we have adopted corresponds to $\lambda=10.$  The  value of $A_{2}$ determines the value of $k_{c}$ and is a free parameter. Most of the results we present are actually independent on the  value of $k_{c}$. Nevertheless $k_c$  determines  the slope of the primordial density spectrum and so affects the detailed results of the bispectra (for non-equilateral configurations).
The bispectra are obtained as in previous section. For the velocity divergence it corresponds to the reduced velocity divergence
$\theta(k)/f_{k}$. We can see that although the modified gravity regime changes the linear growth rate of the density contrast  as soon as $k>k_{c}$, the new couplings terms play a role only for $k\approx k_{c}$ for equilateral type configurations (as illustrated in Figs. \ref{Bispectra}).

\begin{figure}[htbp]
\epsfig{file=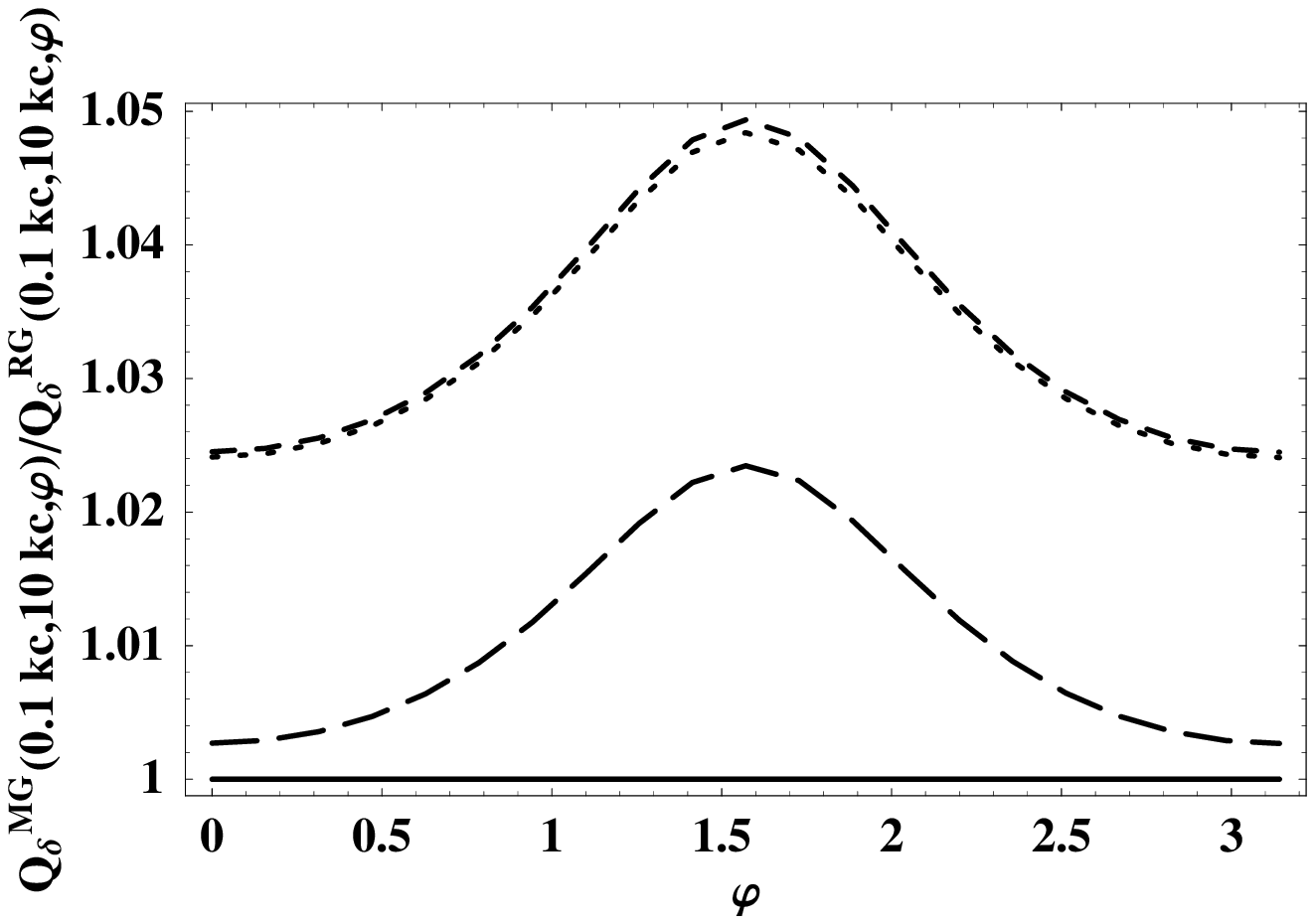,width=6cm}\hspace{.1cm}
\epsfig{file=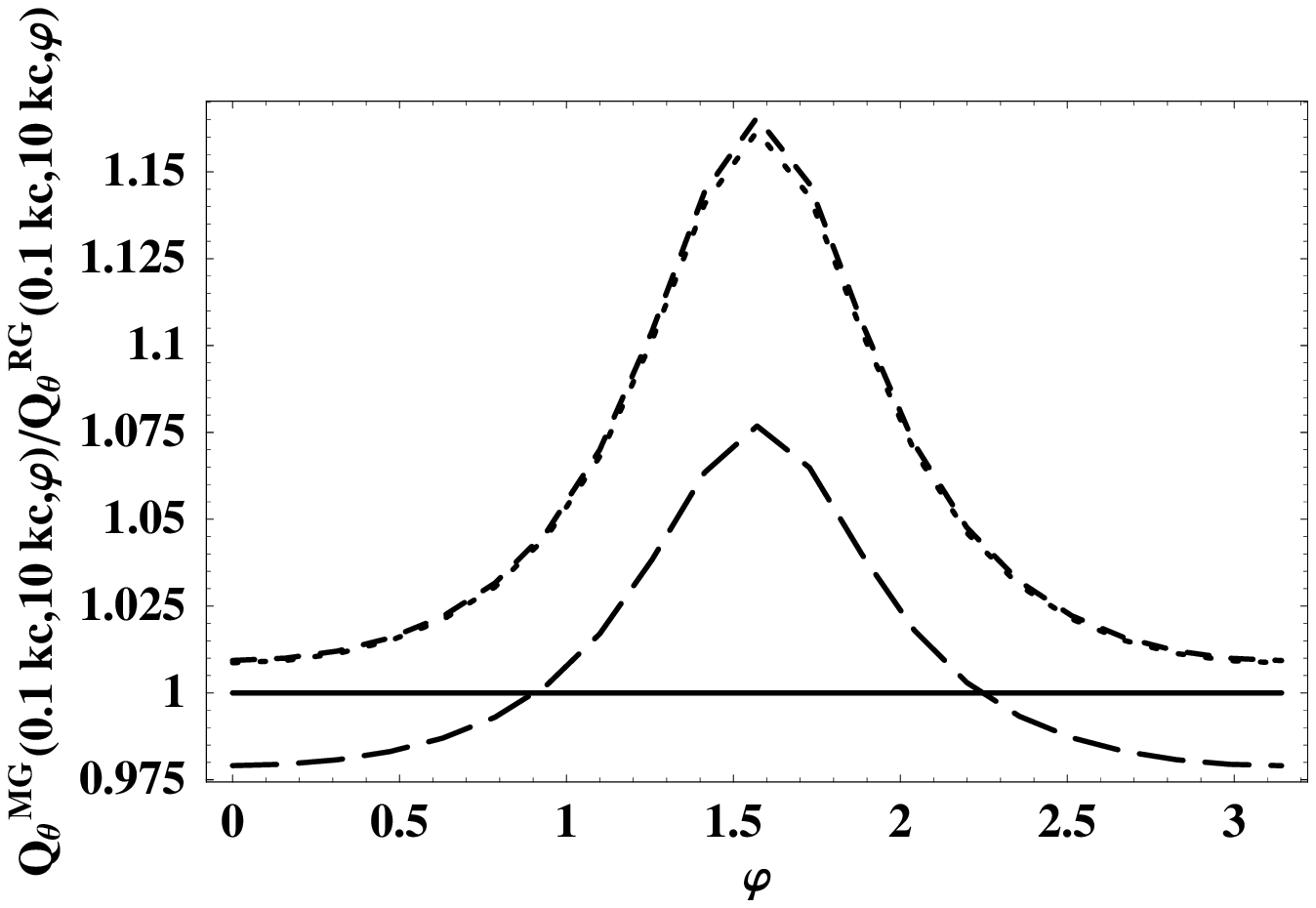,width=6cm}
\caption{The density and reduced velocity divergence bispectra in the squeezed limit for $\lambda=10$. Conventions are the same as for Fig.  \ref{Bispectra}.}
\label{Squeezed}
\end{figure}

In the squeezed limit it is interesting to see that there are significant changes with respect to GR in configurations where
$k_{\rm short}\gg k_{c}$ and $k_{\rm long}\ll k_{c}$ that are due to both changes in the linear growth rates and
to the extra couplings functions. In Fig. \ref{Squeezed} we show the case $k_{\rm short}\approx 10\ k_{c}$ and $k_{\rm long}\approx 0.1 \ k_{c}$.


\section{Conclusion}

We have explored here the consequences of modified gravity models on the evolution of mode coupling amplitudes at large scale.
In this paper we have concentrated our efforts on the computation of the second order expression of the cosmic fluids. For Gaussian initial conditions, the induced kernel is expected to determine the amplitude and shape of the observed bispectra at large enough scale.

We found  that modifications of gravity can change the amplitude of the coupling parameter at different levels. From a pure phenomenological point of view, incorporating a change of gravity into a change of the amplitude of the Euler equation source term
leads to a mild effects in the amplitude of the bispectra. We explicitly computed this effect.
The results are encoded in Eqs. (\ref{nu2fit}-\ref{mu2fit})
in the context of what we called the $\gamma$-model.

Realistic models however lead to a richer phenomenology. The change of gravity is in general both time and scale dependent. This is the case in particular for the chameleon and dilaton models in which the effective strength of gravity is modified through extra scalar degrees
of freedom in which a scale dependent critical wavemode appears which controls the effective gravity strength. It had been stressed
in previous papers that this scale could be of cosmological relevance. We see here that  time and scale dependent effects can play a significant role in the amplitude of the mode couplings for scales below the critical scales.

Furthermore the existence of extra interaction fields lead to a more profound change of the coupling structure. In particular there exist couplings  of arbitrary order, not only quadratic as for standard gravity dynamics at sub-horizon scales (see \cite{2002PhR...367....1B} for details.) This is a situation comparable to that encountered in the DGP models and described in \cite{2009PhRvD..80j4005C,2009PhRvD..80j4006S}. Qualitatively, changing the law of gravity cannot therefore  be captured in a simple change of the linear theory for scales that are comparable with the critical scale. It also implies that the resummation schemes developed in the context of standard gravity (see \cite{2006PhRvD..73f3519C,2006PhRvD..73f3520C,2008PhRvD..78h3503B,2008PhRvD..78j3521B,2010PhRvD..82h3507B}) cannot be directly applied.

\begin{figure}[htbp]
\epsfig{file=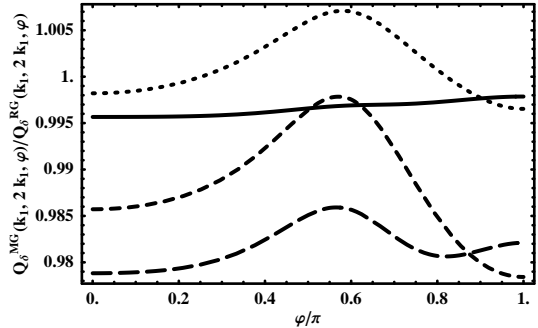,width=7cm}
\caption{The density normalised bispectrum for $k_{2}=2 k_{1}$ and as a function of their relative angle. This is to be compared with
the result presented in \cite{2009PhRvD..80j4005C}.}
\label{ScocciConfigDens}
\end{figure}

Quantitatively we find that the coupling functions $F_{2}$ and $G_{2}$ that describe the 2nd order expression of the cosmic fields can be changed at a few percent level compared to the
standard GR case. The type of changes is however quite model dependent and results in the interplay between transient effects
and direct coupling effects. In Fig. \ref{ScocciConfigDens} we present the relative change in the density bispectrum for a specific set of
configurations. These results are be directly compared to the one found in the context of the DGP model. We can see that
their amplitude is comparable although they differ in detailed scale and shape dependences.

Whether such effects are actually observable in current or future surveys is largely an open question. Measuring
bispectra, with assessed error bars, proves to be  difficult and very little has been done or attempted so far. One example
is to be found in \cite{2001PhRvL..86.1434F} where the bispectrum is measured in the PSCz survey. It has not yet been
measured in the SDSS as such. Only the angular space three-point function has been measured in \cite{2010arXiv1007.2414M}
from which one can infer the error amplitude of such measurements. They are in the percent level range.
This gives us hope that large-scale surveys in preparation can potentially shed lights on these models.

\section*{Acknowledgements}

It is a pleasure to thank F. Vernizzi, E. Sefusatti, J.-Ph. Uzan and L. Hui for  discussions
during the course of this work.
This work was supported the French Agence National de la Recherche under
grant ANR-07-BLAN-0132.

\appendix

\section{The bispectra, full dependence}

In this appendix we simply present the full wavemode dependence of the density and reduced velocity divergence bispectra
when compared to the general relativity case. The modified gravity model we adopted corresponds to $\lambda=10.$
We assume here that the pivot term for $k$ is $k_{c}$ and we assume that at that scale the power spectrum is close to a power law of index $n=-1.5$. We do not expect though that the results are strongly effected by such an hypothesis.

\begin{figure}[htbp]
\epsfig{file=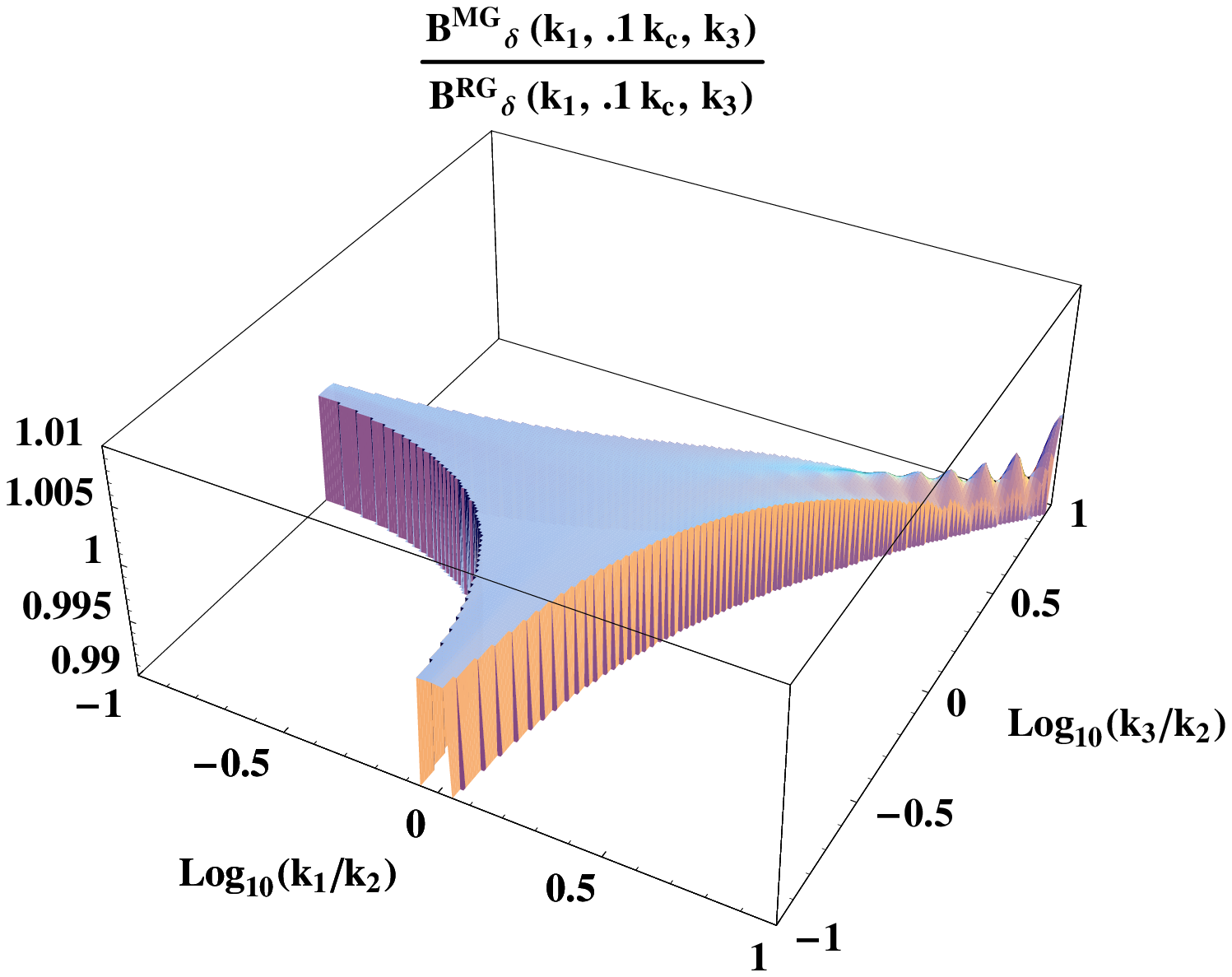,width=7cm}\hspace{.1cm}\epsfig{file=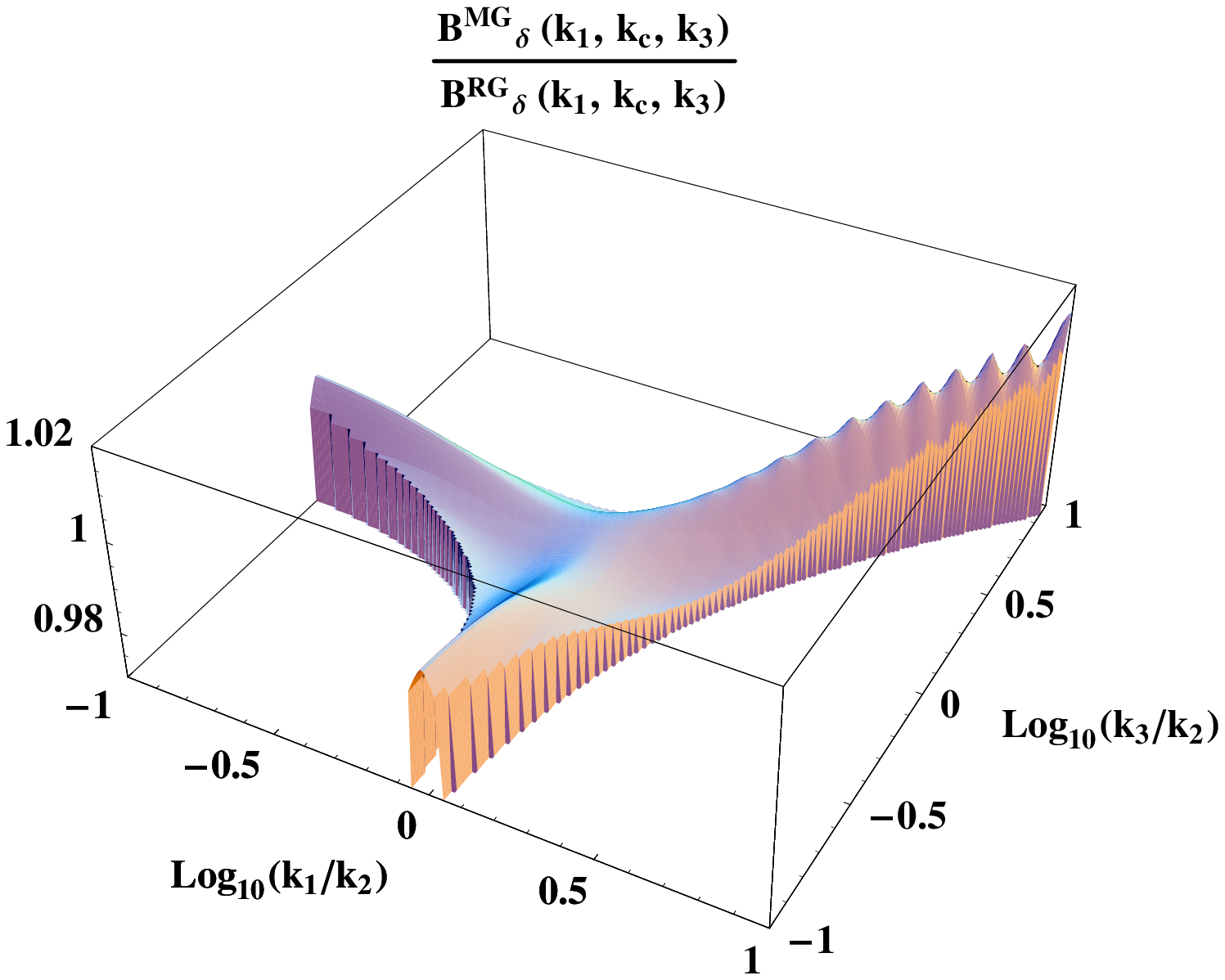,width=7cm}\hspace{.1cm}\epsfig{file=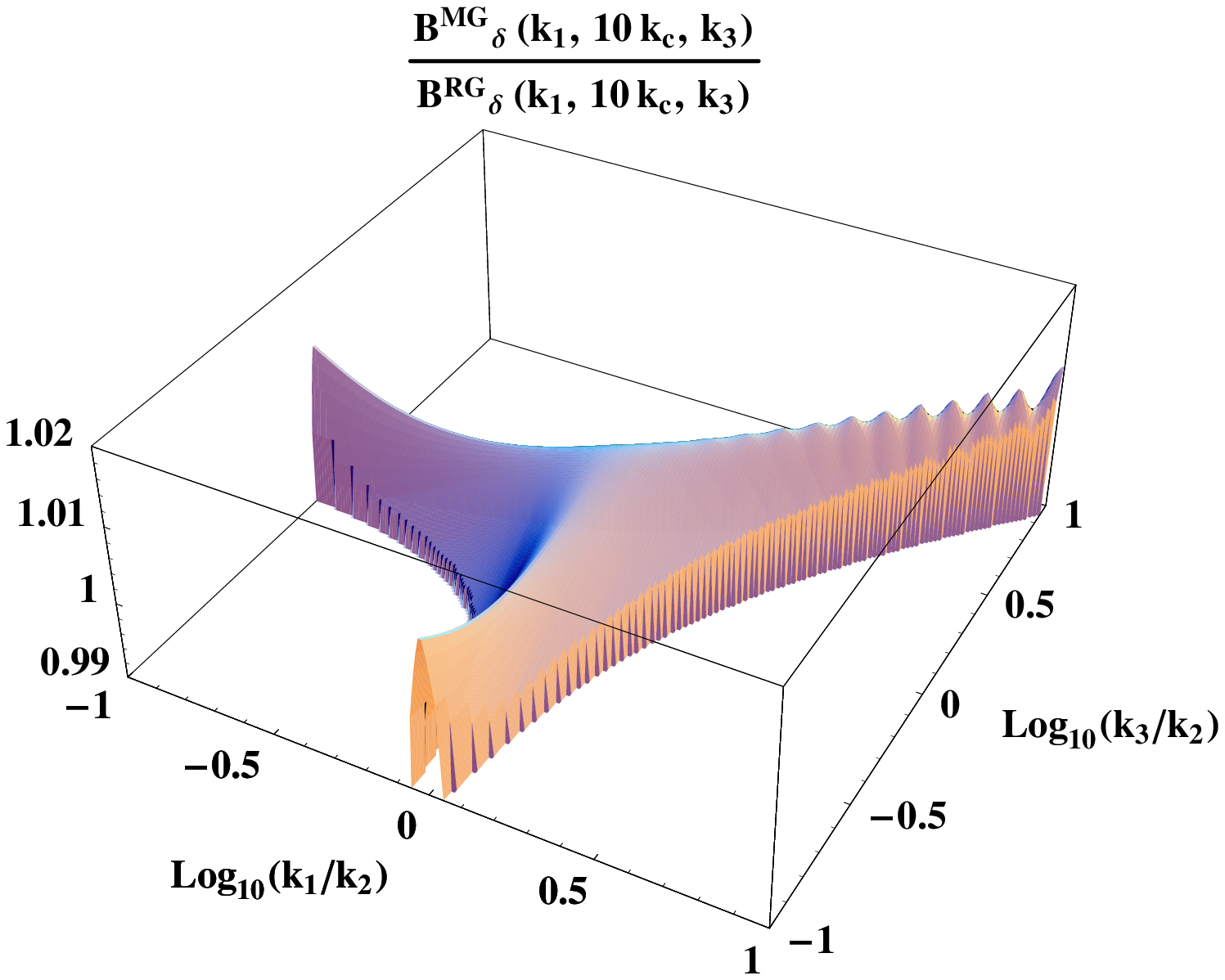,width=7cm}
\caption{Amplitude of the reduced bispectrum $Q_{\delta}(k_{1},k_{2},k_{3})$ for the density field for the modified gravity model ($\lambda=10$) divided by the expected result for standard gravity as a function of $k_{1}/k_{2}$ and $k_{3}/k_{2}$ for respectively $k_{2}=.1 k_{c}$, $k_{2}=k_{c}$ and $k_{2}=10 k_{c}$. We assume here that $P(k)\sim k^{-1.5}.$}
\label{BiSpecD}
\end{figure}

\begin{figure}[htbp]
\epsfig{file=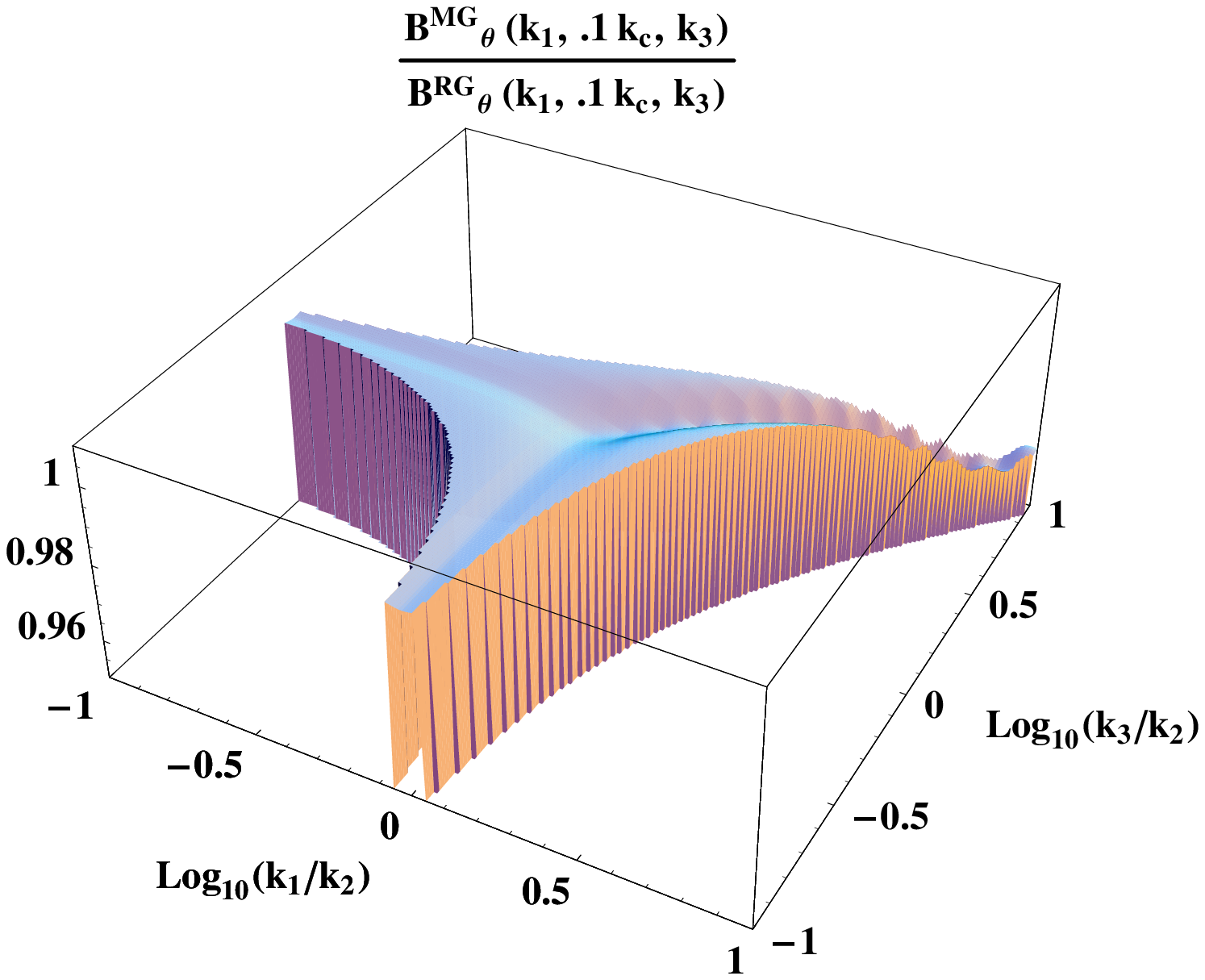,width=7cm}\hspace{.1cm}\epsfig{file=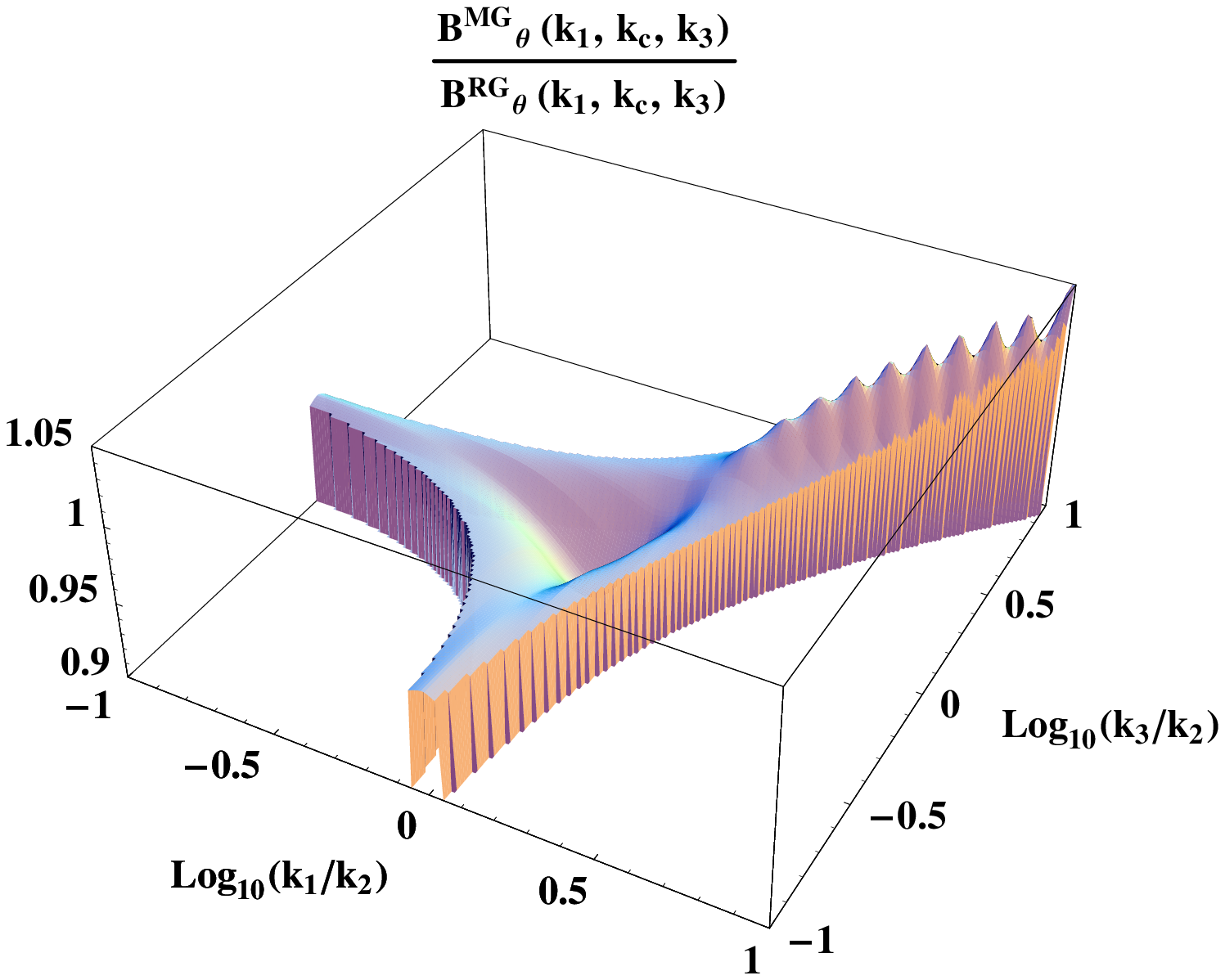,width=7cm}\hspace{.1cm}\epsfig{file=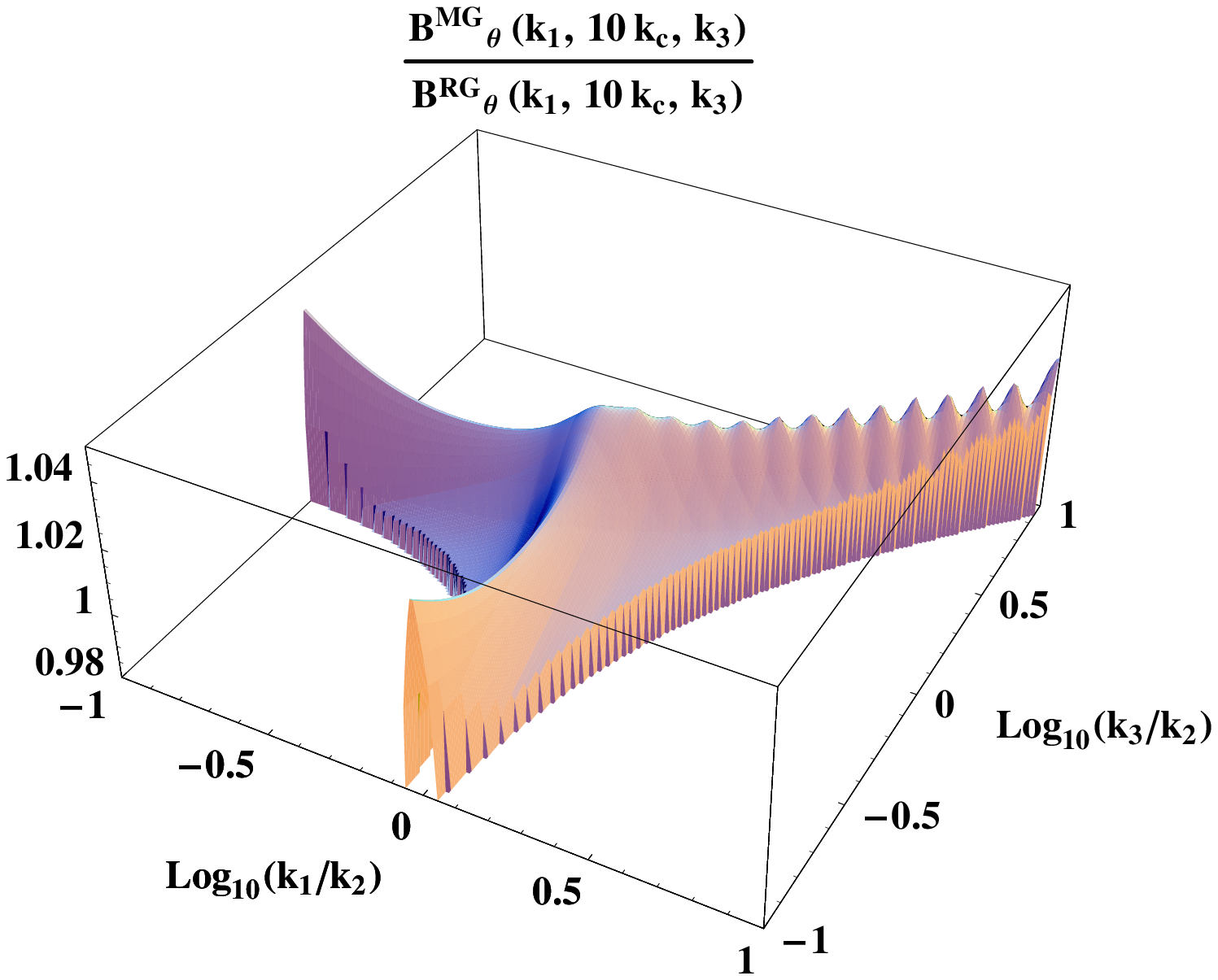,width=7cm}
\caption{Same as previous figure for the reduced velocity divergence.}
\label{BiSpecT}
\end{figure}

The results are presented in Figs. \ref{BiSpecD} and \ref{BiSpecT}
which show the ratio between the modified gravity regime and standard GR. Deviation from GR are observed to be at the percent level
for the density field; up to a few percent level for the reduced velocity field. Note that observed density fields in redshift space correspond to a mixing of these 2 contributions (see for instance \cite{2002PhR...367....1B}).

\ifcqg
\section*{References}
\else
\fi
\ifcqg
\bibliographystyle{h-physrev}
\fi

\bibliography{LSStructure}

\end{document}